\documentclass[reqno,11pt]{amsart}
\usepackage[utf8]{inputenc}
\usepackage{graphicx}
\usepackage{slashed}
\usepackage{amscd}
\usepackage{amssymb}
\usepackage{esint}
\usepackage{enumitem}
\usepackage{comment}
\usepackage{amsmath}
\usepackage{dsfont}
\usepackage[mathscr]{eucal}
\ifx\makepics\undefined
\usepackage[linktocpage]{hyperref}
\usepackage[off]{auto-pst-pdf} 
\else
\usepackage[pspdf=-dALLOWPSTRANSPARENCY]{auto-pst-pdf} 
\fi
\usepackage{pst-all}

\numberwithin{equation}{section}
\allowdisplaybreaks[1]
\definecolor{labelkey}{gray}{.65}

\title[A Canonical Construction of the Extended Hilbert Space]{A Canonical Construction
of the Extended \\ Hilbert Space for Causal Fermion Systems}

\author[F.\ Finster]{Felix Finster}
\address{Fakult\"at f\"ur Mathematik \\ Universit\"at Regensburg \\ D-93040 Regensburg \\ Germany}
\email{finster@ur.de}

\author[P.\ Fischer]{Patrick Fischer \\ \\ April / December 2025}
\address{Fakult\"at f\"ur Mathematik \\ Universit\"at Regensburg \\ D-93040 Regensburg \\ Germany}
\email{patrick.fischer@ur.de}

\newtheorem{Def}{Definition}[section]
\newtheorem{Thm}[Def]{Theorem}
\newtheorem{Prp}[Def]{Proposition}
\newtheorem{Lemma}[Def]{Lemma}
\newtheorem{Remark}[Def]{Remark}

\newcommand{\Thanks}{\vspace*{.5em} \noindent \thanks}
\newcommand{\beq}{\begin{equation}}
\newcommand{\eeq}{\end{equation}}
\newcommand{\Proof}{\begin{proof}}
	\newcommand{\QED}{\end{proof} \noindent}
\newcommand{\QEDrem}{\ \hfill $\Diamond$}

\newcommand{\la}{\langle}
\newcommand{\ra}{\rangle}
\newcommand{\lla}{\langle\!\langle}
\newcommand{\rra}{\rangle\!\rangle}

\newcommand{\Sl}{\mbox{$\prec \!\!$ \nolinebreak}}
\newcommand{\Sr}{\mbox{\nolinebreak $\succ$}}

\newcommand{\C}{\mathbb{C}}
\newcommand{\R}{\mathbb{R}}
\newcommand{\1}{\mathds{1}}

\newcommand{\N}{\mathbb{N}}
\newcommand{\Pdd}{\mbox{$\partial$ \hspace{-1.2 em} $/$}}

\renewcommand{\Tr}{\text{\rm{Tr}}}
\DeclareMathOperator{\tr}{tr}
\renewcommand{\O}{{\mathscr{O}}}
\renewcommand{\L}{{\mathcal{L}}}
\newcommand{\Sact}{{\mathcal{S}}}
\newcommand{\lfe}{\text{\rm{lfe}}}
\newcommand{\Ract}{{\mathcal{R}}}

\newcommand{\eff}{\text{\rm{eff}}}

\renewcommand{\H}{\mathscr{H}}

\newcommand{\Lin}{\text{\rm{L}}}
\newcommand{\F}{{\mathscr{F}}}

\newcommand{\K}{{\mathscr{K}}}

\newcommand{\D}{\mathscr{D}}
\DeclareMathOperator{\re}{Re}
\DeclareMathOperator{\im}{Im}

\DeclareMathOperator{\supp}{supp}

\newcommand{\scrU}{{\mathscr{U}}}

\newcommand{\s}{\mathfrak{s}}

\newcommand{\bu}{{\mathbf{u}}}

\newcommand{\bitem}{\begin{itemize}[leftmargin=1.5em]}
\newcommand{\eitem}{\end{itemize}}

\renewcommand{\sc}{\text{\rm{sc}}}

\newcommand{\x}{\mathbf{x}}
\newcommand{\y}{\mathbf{y}}

\newcommand{\fermi}{{\mathrm{{f}}}}

\newcommand{\Comm}{{\mathscr{C}}}
\newcommand{\scrA}{{\mathscr{A}}}

\DeclareFontFamily{OT1}{rsfso}{}
\DeclareFontShape{OT1}{rsfso}{m}{n}{ <-7> rsfso5 <7-10> rsfso7 <10-> rsfso10}{}
\DeclareMathAlphabet{\mycal}{OT1}{rsfso}{m}{n}

\begin{document}

\begin{abstract}
It is shown that second variations of the causal action can be decomposed into a sum of three terms, two of which being positive and one being small. This gives rise to an approximate decoupling of the linearized field equations into the dynamical wave equation and bosonic field equations. A concrete construction of homogeneous and inhomogeneous solutions of the dynamical wave equation in time strips is presented. In addition, it is show that the solution space admits a positive definite inner product which is preserved under the time evolution. Based on these findings, a canonical construction of the extended Hilbert space containing these solutions is given.
\end{abstract}

\maketitle

\tableofcontents

\section{Introduction} \label{secintro}
The theory of {\em{causal fermion systems}} is a recent approach to fundamental physics
(see the basics in Section~\ref{secprelim}, the textbooks~\cite{cfs, intro}
or the website~\cite{cfsweblink}).
In this approach, all spacetime structures are encoded in a family of one-particle wave functions,
referred to as the {\em{physical wave functions}}.
The physical equations are formulated via the so-called {\em{causal action principle}},
a nonlinear variational principle where an action~$\Sact$ is minimized under variations of all
physical wave functions.
If one describes the Minkowski vacuum, the physical wave functions can be identified with all
the negative-energy solutions of the Dirac equation. In this way, Dirac's original concept of the
Dirac sea is incorporated. Likewise, a non-interacting system involving
particles and/or anti-particles is described by including Dirac solutions of positive frequency
(giving rise to additional one-particle states describing particles) and/or by removing Dirac solutions of negative energy (giving rise to ``holes in the Dirac sea'' describing anti-particles).
If the interaction as described by the causal action principle is taken into account,
the system is no longer composed of Dirac solutions (neither in the vacuum nor in the presence of
classical fields), making it necessary to describe the
whole family of physical wave functions in the realm of many-particle quantum theory
(we refer the interested reader to the recent papers~\cite{fockentangle} and~\cite{fockdynamics}).
Nevertheless, even in the interacting situation, the physical wave functions still give a sensible
notion of ``one-particle wave functions.''

The problem to be addressed in the present paper can be understood already in the non-interacting situation.
In this case, the physical wave functions forming the causal fermion system
span only a proper subspace of the Dirac solution space.
More precisely, in the Minkowski vacuum, the physical wave functions are only the negative-frequency solutions,
but the solutions of positive frequency are not part of the causal fermion system.
This description implements the physical concept that the causal fermion system is formed only of those
wave functions which are ``occupied'' and thus realized in the considered system.
However, this description is not sufficient for formulating the dynamics in terms of
the standard physical equations. In the simplest case, in order to develop the perturbation theory
for Dirac wave functions, one needs Green's operators, whose construction makes it necessary
to know the whole solution space of the Dirac equation (both positive and negative frequencies).
With this in mind, one would like to extend the Hilbert space of physical wave functions
of a causal fermion system to a larger Hilbert space, which for systems in Minkowski space
also includes all solutions of positive frequency.
Indeed, this {\em{extended Hilbert space}} was first constructed in~\cite{dirac},
and it was used in subsequent papers~\cite{fockfermionic, fockentangle, baryogenesis}.

The construction of the extended Hilbert space in~\cite{dirac} has the shortcoming that
it is not manifestly canonical (for details see Appendix~\ref{appdirac}. It is an important
task to overcome this shortcoming and thereby to put the construction of the extended Hilbert space
on a conceptually clear and fully convincing basis.
Compared to the state of knowledge when the paper~\cite{dirac} was written,
there is one new result on the structure of the causal action principle which 
will be very helpful for our constructions: an {\em{approximate decoupling}} of the linearized field equations into
the dynamical wave equation and a bosonic field equation.
This approximate decoupling is based on the observation that second variations of the
causal action principle can be written as a sum of three terms,
the first two being non-negative, and the third being small.
Linearized fields lie in the kernel of the second variations (for details see~\eqref{sum3} and the
explanations thereafter).
Since the first two terms have the same sign, this means that they must both be zero,
except for the coupling described by the third term.
This is what we mean by ``approximate decoupling'' of the linearized field equations.
Clearly, this statement needs to be made precise and formulated in mathematical terms.
This will be worked out in detail in this paper
(Sections~\ref{secsep}--\ref{secdecouple}). Although being a quite remarkable result of independent interest,
the approximate decoupling will serve for us as a preparation for the construction of the
extended Hilbert space (Sections~\ref{secsoldyn} and~\ref{secexhil}).

We now explain a few constructions and ideas in more detail.
The {\em{causal action}}~$\Sact$ has the general structure of a double-integral over spacetime~$M$,
\[ \Sact(\rho) = \iint_{M \times M} \L(x,y)\: d\rho(x)\, d\rho(y) \:, \]
where the Lagrangian~$\L(x,y)$ is a certain nonlocal kernel (basics on causal fermion systems and the
causal action principle will be provided in Section~\ref{seccfsintro} below). 
In the {\em{causal action principle}} one minimizes the causal action under variations
of the integration measure~$\rho$, subject to certain constraints.
A minimizing measure~$\rho$ is a critical point of the causal action in the sense that first variations of the
causal action vanish. This is made precise by the corresponding {\em{Euler-Lagrange (EL) equations}}
(see Section~\ref{secELrestrict}),
which can be thought of as the nonlinear dynamical equations of the theory.
Here we restrict our attention to the {\em{linearized field equations}} obtained by linearizing the
EL equations. More precisely, solutions of the linearized field equations correspond to
first variations which preserve criticality, meaning that
meaning that the corresponding second variations~$\delta^2 \Sact$ of the causal action vanish.
With this in mind, in this paper we always restrict attention to the kernel of~$\delta^2 \Sact$.
The {\em{approximate decoupling}} of the linearized field equations can be understood in
non-technical terms as follows.
We first note that, assuming that the measure~$\rho$ describing the vacuum
is a minimizer of the causal action principle, second variations of the causal action are always non-negative
(for details and consequences see~\cite{positive}). This positivity result, however, will not
sufficient for our purposes. Instead, our argument is based on the observation that the second variations~$\delta^2 \Sact$
can be written as a sum of three terms (for details see Proposition~\ref{prpsecond},
\beq \label{sum3}
\delta^2 \Sact = \delta^2 \Sact^\lfe + \delta^2 \Sact^Q + \Ract \:.
\eeq
the first two of which are non-negative, whereas the third summand (the ``remainder'') is very small.
The positivity of the first summand follows directly from the structure of the causal Lagrangian,
which involves sums of positive terms squared (for details see~\eqref{Lagrange}),
\[ \frac{1}{4n} \sum_{i,j=1}^{2n} \Big( \big|\lambda^{xy}_i \big|
- \big|\lambda^{xy}_j \big| \Big)^2 \:. \]
If each of the two factors of the square is varied linearly, we obtain the
non-negative term
\beq \label{del2Llfeintro}
\delta^2 \L^\lfe(x,y) := 
\frac{1}{4n} \sum_{i,j=1}^{2n} \delta \Big( \big|\lambda^{xy}_i \big|
- \big|\lambda^{xy}_j \big| \Big)\: \delta \Big( \big|\lambda^{xy}_i \big|
- \big|\lambda^{xy}_j \big| \Big) \geq 0 \:,
\eeq
and integrating~$x$ and~$y$ over spacetime gives a positive functional~$\delta^2 \Sact^\lfe \geq 0$.

The positivity of~$\delta^2 \Sact^Q$ is less obvious. This structural property of the causal action
was observed only recently, based on the constructions in~\cite{nonlocal}.
It provides the main new insight compared to the earlier constructions in~\cite{dirac}.
Presenting this positivity property and working out some of its consequences is one of the main objectives
of the present paper. The starting point is to consider a first variation~$\delta \psi$ of one
physical wave function~$\psi$. The corresponding first variation of the causal Lagrangian can be written as
\beq \label{delLintro}
\delta \L(x,y) = -2 \re \Big( \Sl \delta \psi(x)\:  Q(x,y)\, \psi(y) \Sr + \Sl \psi(x)\:  Q(x,y)\, \delta \psi(y) \Sr  \Big)
\eeq
(see~\cite[eq.~(3.8)]{nonlocal} or~\eqref{delLdef} in Section~\ref{secELrestrict}; for details we also refer again
to Section~\ref{secprelim}). This formula can be understood immediately from the fact that~$\delta \L(x,y)$
is real and that it is real-linear in both~$\psi$ and~$\delta \psi$.
The main point is that the kernel~$Q(x,y)$ is zero unless~$x$ and~$y$ are close together.
Therefore, the first variation~$\delta L(x,y)$ vanishes if the supports of~$\psi$ and its variation~$\delta \psi$
are vary far apart. We refer to such variations as {\em{variations with separated supports}}.
They will be introduced systematically in Section~\ref{secsep}, where we also clarify the notions
of ``close together'' and ``far apart'' by discussing the relevant length scales.
Varying once again, we again get zero, unless the second variation
changes the factors~$\psi$ in~\eqref{delLintro} to~$\delta \psi$. We conclude that, for second variations with separated supports,
\beq \label{del2LQintro}
\delta^2 \L(x,y) = -4 \re \Sl \delta \psi(x)\:  Q(x,y)\, \delta \psi(y) \Sr \:.
\eeq
Integrating over~$x$ and~$y$ and using that second variations of the causal action are always
non-negative (following the arguments in~\cite{positive}), we conclude that
\beq \label{del2SQ}
0 \leq \delta^2 \Sact^Q := -4 \int_M d\rho(x) \int_M d\rho(y)\: \Sl \delta \psi(x)\:  Q(x,y)\, \delta \psi(y) \Sr
\eeq
(here the real part could be omitted because taking the complex conjugate corresponds
to interchanging the integration variables~$x \leftrightarrow y$).
This positivity statement holds for any choice of~$\delta \psi$, showing that the
corresponding summand in~\eqref{sum3} is also non-negative for general variations.

All contributions to the second variation which are neither of the form~\eqref{del2Llfeintro} nor of
the form~\eqref{del2LQintro} are subsumed in the remainder term~$\Ract$ in~\eqref{sum3}.
These contributions are small as a consequence of the fact that they contain
either an unvaried factors~$( |\lambda^{xy}_i| - |\lambda^{xy}_j|)$ or else
a factor~$\kappa$ (coming from the second summand of the causal Lagrangian~\eqref{Lagrange}).
Here ``small'' means that, for systems describing Minkowski space, the resulting contributions
to the EL equations will be of higher order in the Planck length
(this will be worked out in detail in Section~\ref{seccouple}).

Using that, for a minimizer, first variations of the form~\eqref{delLintro} must vanish gives
rise to the {\em{restricted EL equations}}
\beq \label{ELintro}
\int_M Q(x,y)\: \psi(y)\: d\rho(y) = \mathfrak{r}\, \psi(x)
\eeq
(for details see~\cite[Section~3.1]{nonlocal} and Section~\ref{secELrestrict};
here the parameter~$\mathfrak{r}$ is the Lagrange parameter of the so-called trace constraint).
This is a linear equation in~$\psi$, which can be regarded as generalizing the Dirac equation to the
setting of causal fermion systems.
Our goal is to extend this equation to more general wave functions which are not necessarily realized in our
system. Also extending the scalar product, this will give the desired extended
Hilbert space~$(\H^\fermi_\rho, \la .|. \ra_\rho)$.

In~\cite{dirac} the extended Hilbert space was constructed by linearly perturbing the kernel~$Q(x,y)$ in~\eqref{ELintro} and analyzing how the solutions change.
A-priori, there is a lot of freedom to choose the linear perturbations.
On the other hand, the additional solutions generated by the perturbations must be compatible
with the conservation laws. This leads to the compatibility conditions which, as already mentioned above,
are not canonical.
A subtle point of this construction, which was not taken into account in the analysis in~\cite{dirac},
is that the considered linear perturbations might come from the summand~$\delta^2 \Sact^\lfe$ in~\eqref{sum3}.
In other words, the EL equations might hold only because contributions from the first and second summands~\eqref{ELintro} cancel each other. However, such cancellations cannot occur
in view of our approximate decoupling, because both contributions necessarily have the same sign.
In order to avoid this contradiction, one must only consider perturbations which do not change
the summand~$\delta^2 \Sact^\lfe$. Unfortunately, it seems difficult to determine this class of
allowed perturbations. With this in mind, the constructions in~\cite{dirac} are not wrong, but incomplete
and not satisfying in their present form.

The general lesson from the approximate decoupling is that the
EL equations have a much more rigid structure than previously thought.
As will be worked out in the present paper, using these additional structural properties of the
minimizing measure and the EL equations, it becomes even easier to construct solutions
of the dynamical wave equation. Our main conclusion is that it is no longer necessary to
perturb the kernel~$Q(x,y)$ in order to obtain the extended solutions.
Instead, the extended Hilbert space is obtained already by solving the corresponding inhomogeneous
solutions in a time strip, with the inhomogeneity being supported near the boundaries of the time strip
in the future and past. More precisely, we consider a causal fermion system which admits
a {\em{global time function}}~$T : M \rightarrow \R$ (for details see Section~\ref{secapprox}).
Given a compact interval~$I$, by the corresponding {\em{time strip}} we mean the spacetime
region~$\Omega_I := T^{-1}(I) \subset M$. The {\em{dynamical wave equation}} is the integral
equation~\eqref{ELintro}, now considered as a linear equation for a general wave function~$\psi$ in spacetime.
We construct solutions of the inhomogeneous dynamical wave equation by inverting the
integral operator (Section~\ref{secinhom}). We show that every homogeneous solution in a slightly smaller time strip
can be obtained with this procedure by a suitable choice of the inhomogeneity support outside the
smaller time strip (Section~\ref{sechom}).
In order to gain further insight into the obtained solutions, we evaluate their
commutator inner product defined by
\[ 
\la \psi | \phi \ra^t := -2i \,\bigg( \int_{\Omega^t} \!d\rho(x) \int_{M \setminus \Omega^t} \!\!\!\!\!\!\!\!d\rho(y) 
- \int_{M \setminus \Omega^t} \!\!\!\!\!\!\!\!d\rho(x) \int_{\Omega^t} \!d\rho(y) \bigg)\:
\Sl \psi(x) \:|\: Q(x,y)\, \phi(y) \Sr_x \:, \]
where~$\Omega^t := T^{-1}(-\infty, t)$ is the past of the time~$T=t$
(for basics see Section~\ref{secconserve} of the preliminaries).
This inner product is conserved (i.e.\ time independent) for homogeneous solutions.
However, it is not conserved for our inhomogeneous solutions, and one can express the
commutator inner product in terms of the inhomogeneity.
We thus conclude that the commutator inner product is in general not positive definite,
but positivity can be arranged by restricting attention to specific inhomogeneities in the past
(Section~\ref{secpositive}). 
Thinking of the cosmological situation, this can be interpreted that
the positivity of the quantum mechanical scalar product is a consequence of boundary conditions 
imposed in the early universe (see Remark~\ref{rembigbang}).
Having constructed a space of solutions endowed with a scalar product, the
extended Hilbert space can be introduced as its completion (see Theorem~\ref{thmmain} in
Section~\ref{secexhil}). Moreover, the conservation of the commutator inner product gives rise to a unitary
time evolution in the extended Hilbert space.

The paper is organized as follows. After providing the necessary background on causal fermion systems,
the causal action principle, the EL equations and the linearized field equations (Section~\ref{secprelim}),
in Section~\ref{secsep} second variations with separated supports are considered.
The positivity statement~\eqref{del2SQ} is proven in Proposition~\ref{prpQpos}.
In Section~\ref{secgenvar} general second variations are considered,
and the decomposition~\eqref{sum3} into two positive terms and a remainder is
derived (Proposition~\ref{prpsecond}).
In Section~\ref{secdecouple} it is worked out that the approximate decoupling gives rise to
estimates for solutions of the linearized field equations in time strips.
These estimates justify the procedure of constructing solutions of the dynamical wave equations
in time strips worked out in Section~\ref{secsoldyn}. The commutator inner product
and its positivity properties are studied in Section~\ref{secpositive}.
Combining these results we can construct the extended Hilbert space
as well as its unitary time evolution (Section~\ref{secexhil}).
We finally explain how the coupling of the dynamical wave equation to the
first and last summand in~\eqref{sum3} can be treated perturbatively, again in a given time strip
(Section~\ref{seccouple}).
in Appendix~\ref{appex} we work out the example of a regularized Dirac dynamics.
This has the purpose of illustrating the solution space, the conservation law of the commutator inner product
and its positivity properties in a way similar as found in the general constructions of this paper.
Finally, Appendix~\ref{appsupplement} provides supplementary material, where
we revisit a few earlier constructions and give a a few preparatory considerations,
with the goal of motivating our constructions better and clarifying the context
in which they were developed.

\section{Preliminaries} \label{secprelim}
We now give the necessary background on causal fermion systems and the causal
action principle. We also introduce the main objects to be used later on.
Our presentation is brief; more details can be found for example in the textbooks~\cite{cfs,intro}.
For the general physical context we recommend Chapter~5 in the recent textbook~\cite{intro}
or the reviews~\cite{dice2014, review}.

\subsection{Causal Fermion Systems and the Reduced Causal Action Principle} \label{seccfsintro}
We begin with the basic definitions.
\begin{Def} \label{defcfs} (causal fermion system) {\em{ 
Given a separable complex Hilbert space~$\H$ with scalar product~$\la .|. \ra_\H$,
we denote the Banach space of bounded linear operators on~$\H$ by~$\Lin(\H)$.
Given a parameter~$n \in \N$ (the {\em{``spin dimension''}}), we let~$\F \subset \Lin(\H)$ be the set of all
symmetric operators on~$\H$ of finite rank, which (counting multiplicities) have
at most~$n$ positive and at most~$n$ negative eigenvalues. On~$\F$ we are given
a positive measure~$\rho$ (defined on a $\sigma$-algebra of subsets of~$\F$).
We refer to~$(\H, \F, \rho)$ as a {\em{causal fermion system}}.
}}
\end{Def} \noindent

A causal fermion system describes a spacetime together
with all structures and objects therein.
In order to single out the physically admissible
causal fermion systems, one must formulate physical equations. To this end, we impose that
the measure~$\rho$ should be a minimizer of the causal action principle,
which we now introduce. For brevity of the presentation, we only consider the
{\em{reduced causal action principle}} where the so-called boundedness constraint has been
incorporated by a Lagrange multiplier term. This simplification does not result in any loss of generality, because
the resulting EL equations are the same as for the non-reduced action principle
as introduced for example in~\cite[Section~\S1.1.1]{cfs}.

For any~$x, y \in \F$, the product~$x y$ is an operator of rank at most~$2n$. 
However, in general it is no longer a symmetric operator because~$(xy)^* = yx$,
and this is different from~$xy$ unless~$x$ and~$y$ commute.
As a consequence, the eigenvalues of the operator~$xy$ are in general complex.
We denote these eigenvalues counting algebraic multiplicities
by~$\lambda^{xy}_1, \ldots, \lambda^{xy}_{2n} \in \C$
(more specifically,
denoting the rank of~$xy$ by~$k \leq 2n$, we choose~$\lambda^{xy}_1, \ldots, \lambda^{xy}_{k}$ as all
the non-zero eigenvalues and set~$\lambda^{xy}_{k+1}, \ldots, \lambda^{xy}_{2n}=0$).
Given a parameter~$\kappa>0$ (which will be kept fixed throughout this paper),
we introduce the $\kappa$-Lagrangian and the causal action by
\begin{align}
\text{\em{$\kappa$-Lagrangian:}} && \L(x,y) &= 
\frac{1}{4n} \sum_{i,j=1}^{2n} \Big( \big|\lambda^{xy}_i \big|
- \big|\lambda^{xy}_j \big| \Big)^2 + \kappa\: \bigg( \sum_{j=1}^{2n} \big|\lambda^{xy}_j \big| \bigg)^2 \label{Lagrange} \\
\text{\em{causal action:}} && \Sact(\rho) &= \iint_{\F \times \F} \L(x,y)\: d\rho(x)\, d\rho(y) \:. \label{Sdef}
\end{align}
The {\em{reduced causal action principle}} is to minimize~$\Sact$ by varying the measure~$\rho$
under the following constraints,
\begin{align}
\text{\em{volume constraint:}} && \rho(\F) = 1 \quad\;\; \label{volconstraint} \\
\text{\em{trace constraint:}} && \int_\F \tr(x)\: d\rho(x) = 1 \:. \label{trconstraint}
\end{align}
This variational principle is mathematically well-posed if~$\H$ is finite-dimensional.
For the existence theory and the analysis of general properties of minimizing measures
we refer to~\cite{continuum, lagrange} or~\cite[Chapter~12]{intro}.
In the existence theory one varies in the class of regular Borel measures
(with respect to the topology on~$\Lin(\H)$ induced by the operator norm),
and the minimizing measure is again in this class. With this in mind, we always assume
that~$\rho$ is a {\em{regular Borel measure}}.

\subsection{The Physical Wave Functions and the Wave Evaluation Operator} \label{secweo}
In the next sections we introduce those inherent structures of a causal fermion system needed
for our analysis. Let~$\rho$ be a {\em{minimizing}} measure. Defining {\em{spacetime}}~$M$ as the support
of this measure,
\[ 
M := \supp \rho \subset \F \:. \]
the spacetime points are symmetric linear operators on~$\H$.
It is a specific feature of the setup of causal fermion systems that each spacetime point
(which later corresponds to of a point of a usual classical spacetime like Minkowski space)
corresponds to a linear operator. In what follows, we shall always identify a spacetime point with
the corresponding operator, which sometimes we refer to for clarity as the {\em{spacetime point operator}}.
These spacetime point operators contain a lot of information which, if interpreted correctly,
gives rise to spacetime structures like causal and metric structures, spinors
and interacting fields (for details see~\cite[Chapter~1]{cfs}).
Here we restrict attention to those structures needed in what follows.
We begin with a basic notion of causality.

\begin{Def} (causal structure) \label{def2}
{\em{For any~$x, y \in \F$, the product~$x y$ is an operator
of rank at most~$2n$. We denote its non-trivial eigenvalues (counting algebraic multiplicities)
by~$\lambda^{xy}_1, \ldots, \lambda^{xy}_{2n}$.
The points~$x$ and~$y$ are
called {\em{spacelike}} separated if all the~$\lambda^{xy}_j$ have the same absolute value.
They are said to be {\em{timelike}} separated if the~$\lambda^{xy}_j$ are all real and do not all 
have the same absolute value.
In all other cases (i.e.\ if the~$\lambda^{xy}_j$ are not all real and do not all 
have the same absolute value),
the points~$x$ and~$y$ are said to be {\em{lightlike}} separated. }}
\end{Def} \noindent
Restricting the causal structure of~$\F$ to~$M$, we get causal relations in spacetime.

Next, for every~$x \in M$ we define the {\em{spin space}}~$S_xM:= x(\H)$
as the image of the operator~$x$; it is a subspace of~$\H$ of dimension at most~$2n$.
It is endowed with the {\em{spin inner product}} $\Sl .|. \Sr_x$ defined by
\[ 
\Sl u | v \Sr_x = -\la u | x v \ra_\H \qquad \text{(for all $u,v \in S_xM$)}\:. \]
A {\em{wave function}}~$\psi$ is defined as a function
which to every~$x \in M$ associates a vector of the corresponding spin space,
\beq \label{psirep}
\psi \::\: M \rightarrow \H \qquad \text{with} \qquad \psi(x) \in S_xM \quad \text{for all~$x \in M$}\:.
\eeq
We remark that a wave function~$\psi$ is said to be {\em{continuous}} if
for every~$x \in M$ and~$\varepsilon>0$ there is~$\delta>0$ such that
\beq \label{wavecontinuous}
\big\| \sqrt{|y|} \,\psi(y) -  \sqrt{|x|}\, \psi(x) \big\|_\H < \varepsilon
\qquad \text{for all~$y \in M$ with~$\|y-x\| \leq \delta$}
\eeq
(where~$|x|$ is the absolute value of the symmetric operator~$x$ on~$\H$, and~$\sqrt{|x|}$
is the square root thereof).
We denote the set of continuous wave functions by~$C^0(M, SM)$.

It is an important observation that every vector~$u \in \H$ of the Hilbert space gives rise to a distinguished
wave function. In order to obtain this wave function, denoted by~$\psi^u$, we simply project the vector~$u$
to the corresponding spin spaces,
\beq \label{psiudef}
\psi^u \::\: M \rightarrow \H\:,\qquad \psi^u(x) = \pi_x u \in S_xM \:,
\eeq
where~$\pi_x : \H \rightarrow S_xM \subset \H$ denotes the orthogonal projection operator.
We refer to~$\psi^u$ as the {\em{physical wave function}} of~$u \in \H$.
A direct computation shows that the physical wave functions are continuous
(in the sense~\eqref{wavecontinuous}). Associating to every vector~$u \in \H$
the corresponding physical wave function gives rise to the {\em{wave evaluation operator}}
\[ 
\Psi \::\: \H \rightarrow C^0(M, SM)\:, \qquad u \mapsto \psi^u \:. \]
Evaluating at a fixed space-time point gives a mapping denoted by
\[ \Psi(x) \::\: \H \rightarrow S_xM\:, \qquad u \mapsto \psi^u(x) \:. \]
Every~$x \in M$ can be written as (for the derivation see~\cite[Lemma~1.1.3]{cfs})
\beq
x = - \Psi(x)^* \,\Psi(x) \label{Fid} \:.
\eeq
In words, every spacetime point operator is the {\em{local correlation operator}} of the wave evaluation operator
at this point (for details see~\cite[\S1.1.4 and Section~1.2]{cfs}).

\subsection{The Restricted Euler-Lagrange Equations} \label{secELrestrict}
We now state the Euler-Lagrange equations.
\begin{Prp} Let~$\rho$ be a minimizer of the reduced causal action principle.
Then the local trace is constant in spacetime, meaning that
\[ 
\tr(x) = 1 \qquad \text{for all~$x \in M$} \:. \]
Moreover, there are parameters~$\mathfrak{r}, \s>0$ such that
the function~$\ell$ defined by
\beq \label{elldef}
\ell \::\: \F \rightarrow \R\:,\qquad \ell(x) := \int_M \L(x,y)\: d\rho(y) - \mathfrak{r}\, \big( \tr(x) -1 \big) - \s
\eeq
is minimal and vanishes in spacetime, i.e.\
\beq \label{EL}
\ell|_M \equiv \inf_\F \ell = 0 \:.
\eeq
\end{Prp} \noindent
For the proof of the EL equations and more details we refer for example to~\cite{nonlocal}.
The parameter~$\mathfrak{r}$ can be viewed as the Lagrange parameter corresponding to the
trace constraint. Likewise, $\s$ is the Lagrange parameter of
the volume constraint.

We now work out what the EL equations mean for first variations of the spacetime points.
The starting point of our consideration is the formula~\eqref{Fid}, which expresses the spacetime point operator
as a local correlation operator. Using this formula, first variations of the wave evaluation operator~$\Psi(x)$ at
a given spacetime point~$x \in M$ give rise to corresponding variations of the spacetime point operator, i.e.\
\beq \label{ufermi}
\bu := \delta x = -\delta \Psi(x)^*\, \Psi(x) - \Psi(x)^*\, \delta \Psi(x) \:.
\eeq
The operator~$\bu$ can be regarded geometrically as a tangent vector to~$\F$ at~$x$.
The minimality of~$\ell$ on~$M$ as expressed by~\eqref{EL} implies that the
derivative of~$\ell$ in the direction of~$\bu$ vanishes, i.e.\
\beq \label{Dul}
D_\bu \ell(x) = 0
\eeq
for all variations of the form~\eqref{ufermi} for which the directional derivative
in~\eqref{Dul} exists. The resulting equations are also referred to as the {\em{restricted EL equations}}.

For the computations, it is more convenient to reformulate
the restricted EL equations in terms of variations of the kernel of the fermionic projector, as we now explain.
In preparation, we use~\eqref{elldef} in order to write~\eqref{Dul} as
\beq \label{resELL}
\int_M D_{1,\bu} \L(x,y)\: d\rho(y) = \mathfrak{r}\, D_\bu \tr(x) \:,
\eeq
where the index one means that the directional derivative acts on the first argument of the Lagrangian.
For the computation of the first variation of the Lagrangian, one can make use of the fact
that for any $p \times q$-matrix~$A$ and any~$q \times p$-matrix~$B$,
the matrix products~$AB$ and~$BA$ have the same non-zero eigenvalues, with the same
algebraic multiplicities. As a consequence, applying again~\eqref{Fid}, we have
\beq
x y 
= \Psi(x)^* \,\big( \Psi(x)\, \Psi(y)^* \Psi(y) \big)
\simeq \big( \Psi(x)\, \Psi(y)^* \Psi(y) \big)\,\Psi(x)^* \:, \label{isospectral}
\eeq
where $\simeq$ means that the operators are isospectral (in the sense that they
have the same non-trivial eigenvalues with the same algebraic multiplicities). Thus, introducing
the {\em{kernel of the fermionic projector}} $P(x,y)$ by
\[ 
P(x,y) := -\Psi(x)\, \Psi(y)^* \::\: S_yM \rightarrow S_xM \:, \]
we can write~\eqref{isospectral} as
\[ x y \simeq P(x,y)\, P(y,x) \::\: S_xM \rightarrow S_xM \:. \]
In this way, the eigenvalues of the operator product~$xy$ as needed for the computation of
the Lagrangian~\eqref{Lagrange} are recovered as
the eigenvalues of a $2n \times 2n$-matrix. Since~$P(y,x) = P(x,y)^*$,
the Lagrangian~$\L(x,y)$ in~\eqref{Lagrange} can be expressed in terms of the kernel~$P(x,y)$.
Consequently, the first variation of the Lagrangian can be expressed in terms
of the first variation of this kernel. Being real-valued and real-linear in~$\delta P(x,y)$,
it can be written as
\beq \label{delLdef}
\delta \L(x,y) = 2 \re \Tr_{S_xM} \!\big( Q(x,y)\, \delta P(x,y)^* \big)
\eeq
(where~$\Tr_{S_xM}$ denotes the trace on the spin space~$S_xM$)
with a kernel~$Q(x,y)$ which is again symmetric (with respect to the spin inner product), i.e.\
\beq \label{Qxydef}
Q(x,y) \::\: S_yM \rightarrow S_xM \qquad \text{and} \qquad Q(x,y)^* = Q(y,x) \:.
\eeq
More details on this method and many computations can be found in~\cite[Sections~1.4 and~2.6
as well as Chapters~3-5]{cfs}. Expressing the variation of~$P(x,y)$ in terms of~$\delta \Psi$,
the first variations of the Lagrangian can be written as
\begin{align*}
D_{1,\bu} \L(x,y) = 2\, \re \tr \big( \delta \Psi(x)^* \, Q(x,y)\, \Psi(y) \big) \\
D_{2,\bu} \L(x,y) = 2\, \re \tr \big( \Psi(x)^* \, Q(x,y)\, \delta \Psi(y) \big)
\end{align*}
(where~$\tr$ denotes the trace of a finite-rank operator on~$\H$).
Using these formulas, the restricted EL equation~\eqref{resELL} becomes
\[ \re \int_M \tr \big( \delta \Psi(x)^* \, Q(x,y)\, \Psi(y) \big)\: d\rho(y) = \mathfrak{r}\,\re \tr \big( \delta \Psi(x)^* \,\Psi(x) \big)  \:. \]
Using that the variation can be arbitrary at every spacetime point, we obtain
\[ \int_M Q(x,y)\, \Psi(y) \:d\rho(y) = \mathfrak{r}\, \Psi(x) \qquad \text{for all~$x \in M$}\:, \]
where~$\mathfrak{r} \in \R$ is the Lagrange parameter of the trace constraint.
Denoting the integral operator with kernel~$Q(x,y)$ by~$Q$, the
restricted EL equations can be written in the shorter form
\beq \label{ELQ}
Q \Psi = \mathfrak{r}\, \Psi \:.
\eeq

\subsection{The Linearized Field Equations in Wave Charts} \label{seclinfield}
Following the procedure in~\cite[Section~3]{nonlocal}, we describe the linearized field equations
exclusively in wave charts. This has the main advantage that the bosonic and fermionic equations
are described in a unified way. We now briefly explain the resulting formalism
and refer for the detailed derivation to~\cite[Section~3.2]{nonlocal}.

The linearized field equations describe variations of the measure~$\rho$ which preserve the EL equations.
More precisely, we vary the spacetime point operators again
according to~\eqref{ufermi} by varying the wave evaluation operator in~\eqref{Fid}.
In this case, preserving the restricted EL equations~\eqref{ELQ} means that
\beq \label{lfehom}
(DQ|_{\Psi}(\delta \Psi)) \,\Psi + Q\, \delta \Psi - \mathfrak{r}\, \delta \Psi = 0\:,
\eeq
where~$DQ|_{\Psi}(\delta \Psi)$ is the variational derivative of the kernel~$Q(x,y)$ under the first variation
of the wave evaluation operator~$\delta \Psi$.
The equations~\eqref{lfehom} are the {\em{homogeneous linearized field
equations}}. It is useful to allow for an inhomogeneity on the right side of the equations.
Thus we write the {\em{inhomogeneous linearized field equations}} as
\beq \label{lfe}
(DQ|_{\Psi}(\delta \Psi)) \,\Psi + Q\, \delta \Psi - \mathfrak{r}\, \delta \Psi = \Xi \:,
\eeq
where the inhomogeneity is a given mapping
\[ \Xi \::\: \H \rightarrow C^0(M, SM) \:. \]
These equations were formulated and analyzed computationally in~\cite{pfp, cfs}.
In~\cite{nonlocal} they are derived in detail in wave charts, based on~\cite{gaugefix, banach}.
We refer to the resulting equations~\eqref{lfehom} and~\eqref{lfe} as the {\em{linearized field equations in
wave charts}}. They take the form
\[ \int_M \Big( \delta Q^\alpha|_\beta(x,y) \,\Psi^\beta(y) + Q^\alpha|_\beta(x,y)\, \delta \Psi^\beta(y) \Big)\, d\rho(y)
= \mathfrak{r}\, \delta \Psi^\alpha(x) \:, \]
with kernels given by
\begin{align*}
\delta Q^\alpha|_\beta(x,y) &= K^{\alpha \gamma}|_{\beta \delta}(x,y)\, \delta P^\delta|_\gamma(y,x)
+ K^\alpha_{\;\;\,\gamma}|_\beta^{\;\;\,\delta}(x,y)\, \delta P^\gamma|_\delta(x,y) \\
\delta P^\alpha|_\beta(x,y) &= -\big(\delta \Psi \big)^\alpha(x)\, \Psi^*_\beta(y)
-\Psi^\alpha(x)\, \big(\delta \Psi \big)^*_\beta(y) \:. 
\end{align*}

\subsection{The Conserved Commutator Inner Product} \label{secconserve}
The connection between symmetries and conservation laws made by Noether's theorem
extends to causal fermion systems~\cite{noether}. However, the conserved quantities of a causal fermion system
have a rather different structure, being formulated in terms of so-called surface layer integrals.
A {\em{surface layer integral}} is a double integral of the form
\[ 
\int_\Omega \bigg( \int_{M \setminus \Omega} (\cdots)\: \L(x,y)\: d\rho(y) \bigg)\, d\rho(x) \]
where the two variables~$x$ and~$y$ are integrated over~$\Omega$ and its complement,
and~$(\cdots)$ stands for variational derivatives acting on the Lagrangian.
Since in typical applications,
the Lagrangian is small if~$x$ and~$y$ are far apart, the main contribution to the
surface layer integral is attained when both~$x$ and~$y$ are near the boundary~$\partial \Omega$.
With this in mind, a surface layer integral can be thought of as a ``thickened'' surface integral,
where we integrate over a spacetime strip of a certain width. For systems in Minkowski space,
constructed of solutions of the Dirac equation of mass~$m$ (for details see for
example~\cite[Section~1.2]{cfs} or~\cite[Section~5.5]{intro}),
the length scale of this strip is the Compton scale~$m^{-1}$.
For more details on the concept of a surface layer integral we refer to~\cite[Section~9.1]{intro}.

There are various Noether-like theorems for causal fermion systems, which relate
symmetries to conservation laws (for an overview 
see~\cite{osi} or~\cite[Chapter~9]{intro}). The conserved quantity of relevance here is the
commutator inner product (for more details see~\cite[Section~9.4]{intro}
or~\cite[Section~5]{noether} and~\cite[Section~3]{dirac}):
The causal action principle is invariant under unitary transformations of the measure~$\rho$, i.e.\ under
transformations
\[ \rho \rightarrow \scrU \rho \qquad \text{with} \qquad
(\scrU \rho)(\Omega) := \rho \big( \scrU^{-1} \,\Omega\, \scrU \big) \:, \]
where~$\scrU$ is a unitary operator on~$\H$ and~$\Omega \subset \F$ is any measurable subset.
The conserved quantity corresponding to this symmetry is the so-called {\em{commutator inner product}}
\beq \label{cip}
\la \psi | \phi \ra^\Omega := -2i \,\bigg( \int_{\Omega} \!d\rho(x) \int_{M \setminus \Omega} \!\!\!\!\!\!\!\!d\rho(y) 
- \int_{M \setminus \Omega} \!\!\!\!\!\!\!\!d\rho(x) \int_{\Omega} \!d\rho(y) \bigg)\:
\Sl \psi(x) \:|\: Q(x,y)\, \phi(y) \Sr_x \:,
\eeq
where~$\psi, \phi$ are wave functions~\eqref{psirep}, and~$\Omega \subset M$ describes a
spacetime region (and the kernel~$Q(x,y)$ as in~\eqref{Qxydef}).
The set~$\Omega$ should be thought of as the past of a Cauchy surface
at time~$t$, so that the surface layer integral describes a ``thickened'' integral over the Cauchy surface.
Then the conservation law states that the commutator inner product of any two physical wave
functions~$\psi$ and~$\phi$ (as defined by~\eqref{psiudef}) does not depend on the choice of
the Cauchy surface. 

We note for clarity that talking of the ``past of a Cauchy surface'' makes it necessary that
spacetime is time-orientable. This and related notions have been made precise in~\cite[Section~3.2]{dirac}.
Here we do not need to enter the details but assume for simplicity that our spacetime admits a global
time function (as was mentioned in the introduction and will be defined in Section~\ref{secapprox}).

One of the questions to be analyzed in what follows is if and under which assumptions the commutator
inner product coincides (up to an irrelevant prefactor) with the Hilbert space scalar product.
The next notion first introduced in~\cite{dirac} makes it possible to relate the two inner products.
\begin{Def} \label{defSLrep}
Let~$\H^\fermi$ be a subspace of the Hilbert space~$(\H, \la .|. \ra_\H)$.
The commutator inner product is said to {\bf{represent the scalar product}} on~$\H^\fermi$ if
\beq \label{Ccond}
\la u|v \ra^\Omega_\rho = c\, \la u|v \ra_\H \qquad \text{for all~$u,v \in \H^\fermi$}
\eeq
with a suitable positive constant~$c$.
\end{Def} \noindent
We point out that not all causal fermion systems have this property.
Concrete counter examples are spacetimes with a non-zero fermionic von Neumann entropy
or with thermal states; see~\cite{fermientropy, desittercfs}.
Moreover, it is shown in Appendix~\ref{appA} that this property cannot hold on the whole
Hilbert space (i.e.\ for~$\H^\fermi = \H$).

\section{Second Variations with Separated Supports} \label{secsep}
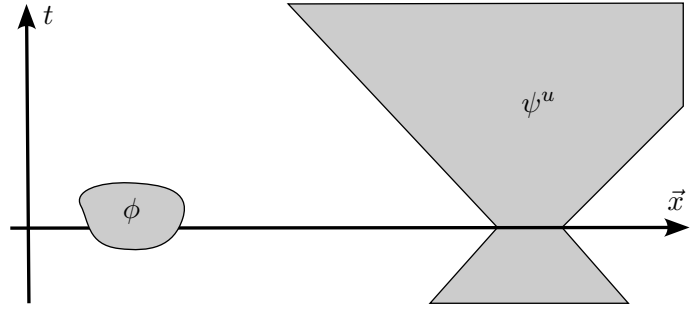
\begin{figure}
\psset{xunit=.5pt,yunit=.5pt,runit=.5pt}
\begin{pspicture}(511.86317744,226.09504279)
{
\newrgbcolor{curcolor}{0.80000001 0.80000001 0.80000001}
\pscustom[linestyle=none,fillstyle=solid,fillcolor=curcolor]
{
\newpath
\moveto(511.8631748,226.09504279)
\lineto(511.8631748,148.42920216)
\lineto(420.59926299,57.08572752)
\lineto(470.49605669,0.00000027)
\lineto(321.23452724,0.00000027)
\lineto(372.01366299,57.08110137)
\lineto(214.42370646,225.44962043)
\closepath
}
}
{
\newrgbcolor{curcolor}{0 0 0}
\pscustom[linewidth=0.99999871,linecolor=curcolor]
{
\newpath
\moveto(511.8631748,226.09504279)
\lineto(511.8631748,148.42920216)
\lineto(420.59926299,57.08572752)
\lineto(470.49605669,0.00000027)
\lineto(321.23452724,0.00000027)
\lineto(372.01366299,57.08110137)
\lineto(214.42370646,225.44962043)
\closepath
}
}
{
\newrgbcolor{curcolor}{0 0 0}
\pscustom[linewidth=2.4999988,linecolor=curcolor]
{
\newpath
\moveto(5.25730772,56.59984657)
\lineto(500.90284724,57.08451429)
}
}
{
\newrgbcolor{curcolor}{0 0 0}
\pscustom[linestyle=none,fillstyle=solid,fillcolor=curcolor]
{
\newpath
\moveto(497.39600565,64.0810851)
\lineto(516.65283214,57.09991543)
\lineto(497.40969555,50.08109852)
\curveto(500.90570501,54.16201709)(500.90000659,59.9895115)(497.39600565,64.0810851)
\closepath
}
}
{
\newrgbcolor{curcolor}{0 0 0}
\pscustom[linewidth=2.4999988,linecolor=curcolor]
{
\newpath
\moveto(19.27655055,0.0101672)
\lineto(17.57166236,209.70014767)
}
}
{
\newrgbcolor{curcolor}{0 0 0}
\pscustom[linestyle=none,fillstyle=solid,fillcolor=curcolor]
{
\newpath
\moveto(10.60035294,206.14335332)
\lineto(17.443611,225.44961955)
\lineto(24.5998835,206.25717675)
\curveto(20.49406437,209.72390832)(14.66675977,209.67652931)(10.60035294,206.14335332)
\closepath
}
}
{
\newrgbcolor{curcolor}{0.80000001 0.80000001 0.80000001}
\pscustom[linestyle=none,fillstyle=solid,fillcolor=curcolor]
{
\newpath
\moveto(95.15506016,90.68583712)
\curveto(71.33802709,91.63975964)(52.4289411,84.35164752)(58.38340157,73.33208342)
\curveto(64.33786205,62.31251933)(62.9888315,44.12990389)(86.8541178,41.23551523)
\curveto(110.71940409,38.34113035)(128.3219263,42.42470578)(135.10154079,64.08110389)
\curveto(141.88115528,85.73750578)(118.97209323,89.73191082)(95.15506016,90.68583712)
\closepath
}
}
{
\newrgbcolor{curcolor}{0 0 0}
\pscustom[linewidth=0.99999871,linecolor=curcolor]
{
\newpath
\moveto(95.15506016,90.68583712)
\curveto(71.33802709,91.63975964)(52.4289411,84.35164752)(58.38340157,73.33208342)
\curveto(64.33786205,62.31251933)(62.9888315,44.12990389)(86.8541178,41.23551523)
\curveto(110.71940409,38.34113035)(128.3219263,42.42470578)(135.10154079,64.08110389)
\curveto(141.88115528,85.73750578)(118.97209323,89.73191082)(95.15506016,90.68583712)
\closepath
}
\rput[bl](30,210){\normalsize{$t$}}
\rput[bl](500,70){\normalsize{$\vec{x}$}}
\rput[bl](390,140){\normalsize{$\psi^u$}}
\rput[bl](90,60){\normalsize{$\phi$}}
}
\end{pspicture}
\caption{A variation with separated supports.}
\label{figseparated}
\end{figure}

Before entering the general analysis of second variations, in this section we shall consider
a special class of variations whose treatment is particularly simple.
In order to describe these variations, we choose~$u \in \H$ and denote the corresponding
physical wave function by~$\psi^u$ (see~\eqref{psiudef}). We vary this physical wave function according to
\beq \label{psiuvary}
\tilde{\psi}^u_\tau := \psi^u + \tau \phi \:,
\eeq
where~$\phi \in C^0_0(M, SM)$ is a compactly supported wave function.
The general idea is to choose~$\phi$ in such a way that the supports of~$\phi$ and~$\psi^u$
are far apart. One specific scenario, where~$\psi^u$ is assumed to have spatially compact support,
is shown in Figure~\ref{figseparated}. 
The assumption of spatially compact support is too strong in view
of Hegerfeldt's theorem for negative-energy wave packets (see for example~\cite{split}).
Therefore, it is preferable to merely assume that~$\psi^u$ is very small in a neighborhood
of the support of~$\phi$. We implement this assumption by the approximation that the first variation
of the kernel of the fermionic projector vanishes in all composite expressions obtained by varying the Lagrangian.
We write this assumption simply as
\beq \label{delPapprox}
\delta P(x,y) = -|\phi(x)\Sr \Sl \psi^u(y)| - |\psi^u(x)\Sr \Sl \phi^u(y)| \approx 0 \:,
\eeq
and understand implicitly that this should hold for all~$x$ and~$y$ for which the Lagrangian~$\L(x,y)$
and its variations are non-zero.
We refer to variations satisfying this condition as {\em{variations with separated supports}}.
However, {\em{second}} variations of~$P(x,y)$ do not vanish, because
\beq \label{del2P}
\delta^2 P(x,y) = -2\, |\phi(x)\Sr \Sl \phi(y)| \:.
\eeq

\begin{Lemma} \label{lemmavary}
For variations~\eqref{psiuvary} satisfying~\eqref{delPapprox},
the first and second variations of the Lagrangian and the local trace are given by
\begin{align}
\delta \L(x,y) &\approx 0 \:,\qquad \delta \tr(x) \approx  0 \\
\delta^2 \L(x,y) &\approx -2 \re \big( \Sl \phi(x)\:|\: Q(x,y)\, \phi(y) \Sr_x \big) \label{d2Lsep} \\
\delta^2 \tr(x) &= -2\,\Sl \phi(x) | \phi(x) \Sr_x \:,
\end{align}
where~$Q(x,y)$ is the kernel describing first variations of the Lagrangian~\eqref{delLdef}
(and the symbol~``$\approx$'' refers again to the assumption~\eqref{delPapprox}).
\end{Lemma}
\Proof Combining the formulas for the first variations~\eqref{delLdef} and
\[ \delta \tr(x) = \Tr_{S_xM} \!\big(  \delta P(x,x) \big) \]
with~\eqref{delPapprox}, one sees that the first variations vanish.
When computing second variations, it suffices to vary the kernels~$\delta P(x,y)$ and~$\delta P(x,x)$.
Using~\eqref{del2P} gives the result.
\QED
Before going on, we comment on the error term of the approximation~\eqref{delPapprox}.
In the general setup considered here, it is not possible to quantify the error term.
But, considering a more specific situation, we can at least explain the scaling of the error terms.
To this end, we consider the a causal fermion system describing our Universe,
obtained from Dirac wave functions in the classical Lorentzian spacetime
(the general construction is outlined in~\cite[Section~1]{nrstg}).
This construction involves the usual physical length scales like the Compton scale and the
age of the Universe.
In a laboratory, the physical wave function~$\psi^u$ can be chosen to be localized
in space on the Compton scale, up to rapidly decaying errors (for details see~\cite{split}).
The spatial distance to the support of~$\phi$, on the other hand,
can be chosen as large as the size of the known Universe.
Therefore, the relative error in~\eqref{delPapprox} can be made as small as
\beq \label{errorscale}
\O \Big( \big( m \times \text{(size of the Universe)} \big)^{-p} \Big) \approx
\O \big( 10^{-39 \,p} \big) \qquad \text{with} \qquad p \in \N\:,
\eeq
where the parameter~$p$ can be chosen arbitrarily large (here we chose the size of the Universe as~$93$
billion light years and~$m$ as the Compton wave length of the electron). One sees that
the error terms are even much smaller than higher order Planck scale corrections. With this in mind,
we will simply disregard the error terms in what follows.

We now work out corresponding second variations of the causal action.
As observed in~\cite{lagrange, nonlocal}, minimizers of the causal action principle
are also obtained by minimizing the {\em{effective action}}~$\Sact^\text{\rm{eff}}$ defined by
\[ \Sact^\text{\rm{eff}}(\rho) := \int_\F d\rho(x) \int_\F d\rho(y)\: \L(x,y) 
- 2 \int_\F \Big( \mathfrak{r}\, \big( \tr (x) -1 \big) + \s \Big) \,d\rho(x) \:. \]
Indeed, considering first variations of the measure~$\rho$, one immediately verifies
that stationarity of the effective action gives the EL equations with~$\ell$ according to~\eqref{elldef},
including the Lagrange multiplier terms. In this sense, both the volume and the trace constraints can be treated
with Lagrange multipliers.
Moreover, since the considered variations~\eqref{psiuvary} clearly respect the volume constraint, we
may simplify the effective action to
\beq \label{Seff}
\Sact^\text{\rm{eff}}(\rho) := \int_\F d\rho(x) \int_\F d\rho(y)\: \L(x,y) 
- 2 \mathfrak{r} \int_\F \tr (x)\,d\rho(x) \:.
\eeq

Since~$\rho$ is a minimizer, second variations are non-negative (for details see~\cite{positive}).
A direct computation using Lemma~\ref{lemmavary} yields the following result.
\begin{Prp}  \label{prpQpos}
For any wave function~$\phi \in C^0_0(M, SM)$,
\[  -4 \int_M \Sl \phi(x) \:|\: \big((Q - \mathfrak{r}) \phi \big)(x) \Sr_x\: d\rho(x) \geq 0 \:, \]
up to errors of the order~\eqref{errorscale}.
\end{Prp}

\section{General Second Variations} \label{secgenvar}
In the previous section, we considered variations of an individual physical wave functions~\eqref{psiuvary}.
Now we generalize this procedure by varying all physical wave functions at the same time.
To this end, we consider general variations of the wave evaluation operator of the form
\beq \label{Psivar}
\tilde{\Psi}_\tau = \Psi + \tau \Phi \qquad \text{with} \qquad \Phi \in \Lin(\H, C^0_0(M, SM))\:.
\eeq
Clearly, these variations comprise the variations with separated supports~\eqref{psiuvary}
and~\eqref{delPapprox} analyzed in the previous section.
We again denote variations in the parameter~$\tau$ by~$\delta$, $\delta^2$, and so on. In particular,
\[ \delta \Psi := \frac{d}{d\tau} \tilde{\Psi}_\tau \Big|_{\tau=0} = \Phi \qquad \text{and} \qquad
\delta^2 \Psi := \frac{d^2}{d\tau^2} \tilde{\Psi}_\tau \Big|_{\tau=0} = 0 \]
and
\[ \delta P(x,y) = -\delta \Psi(x)\, \Psi(y)^* - \Psi(x)\, \delta \Psi(y)^* \:,\qquad
\delta^2 P(x,y) = -\delta \Psi(x)\, \delta \Psi(y)^* \:. \]
In order to ensure that the following computations are mathematically well-defined,
one needs to make suitable assumptions on the mapping~$\Phi$. One possibility is to restrict attention
to smooth and compact variations (see~\cite[Def.~1.4.2]{cfs}).

\subsection{Perturbation Expansion of the Eigenvalues} \label{secperteigen}
The causal action principle is formulated in terms of the eigenvalues~$(\lambda^{xy}_i)_{i=1,\ldots, 4n}$
of the closed chain
\beq \label{Axy}
A_{xy}:= P(x,y)\, P(y,x) \:. 
\eeq
These eigenvalues can be computed perturbatively
(for details see~\cite[Chapter~5 and Appendix~G]{pfp} or~\cite[Section~2.6 and Appendix~B]{cfs}).
We now write a few general structural results of this analysis using a convenient symbolic notation.
For our purposes, it suffices to compute the eigenvalues of the closed chain to second order
in perturbation theory. Decomposing the closed chain as
\[ A_{xy} = A_{xy}^\text{vac} + \Delta A \]
(where~$A_{xy}^\text{vac}$ is the closed chain in the vacuum), we denote the
orders in~$\Delta A$ with a superscript in round brackets, i.e.\
\[ \lambda^{xy}_i = \lambda^{(0)}_i + \lambda^{(1)}_i + \lambda^{(2)}_i + \cdots \:. \]
The first order in $\Delta A$ can be written as
\[ \lambda^{(1)}_i = \Tr_{S_x} \big( \Lambda^{xy}_i \:\Delta A_{xy} \big) \:, \]
where~$\Lambda^{xy}_i$ denotes the spectral projection operators in the vacuum.
Using~\eqref{Axy} together with symmetry property~$P(y,x) = P(x,y)^*$, we have
\begin{align}
\delta \lambda_i &= \lambda^{(1)}_i  + \O \big( (\delta P)^2 \big) \notag \\
&= \Tr_{S_x} \big( \Lambda^{xy}_i \:\delta P(x,y)\, P(y,x) \big)
+ \Tr_{S_x} \big( \Lambda^{xy}_i \:P(x,y)\, \delta P(y,x) \big) \notag \\
&= \Tr_{S_y} \big( \Lambda^{yx}_i \: P(y,x)\, \delta P(x,y) \big)
+ \Tr_{S_x} \big( \Lambda^{xy}_i \:P(x,y)\, \delta P(y,x) \big) \label{dellam1}
\end{align}
(where~$\O( (\delta P)^2)$ means that we disregard all terms being quadratic
in~$\delta P(x,y)$ and/or $\delta P(y,x)$).
Considering second variations, also the first order perturbation of the eigenvalues comes into play,
involving a second variation of the perturbation operator. More precisely,
\begin{align}
\delta^2 \lambda_i &= \Tr_{S_y} \big( \Lambda^{yx}_i \: P(y,x)\, \delta^2 P(x,y) \big) + \Tr_{S_x} \big( \Lambda^{xy}_i \:P(x,y)\, \delta^2 P(y,x) \big) \notag \\
&\quad\: + 2\, \Tr_{S_x} \big( \Lambda^{xy}_i \:\delta P(x,y)\, \delta P(y,x) \big) 
+ \lambda^{(2)}_i + \O \big( (\delta P)^3 \big) \:. \label{dellam2}
\end{align}
Using this formula, the absolute values of the eigenvalues can be perturbed with the help of the chain rule,
\begin{align}
\delta^2 \big|\lambda^{xy}_i \big| &= \delta^2 \sqrt{ \overline{\lambda^{xy}_i} \lambda^{xy}_i } \notag \\
&= \frac{1}{2\, |\lambda^{xy}_i|}\: \Big( \overline{\delta^2 \lambda^{xy}_i}\, \lambda^{xy}_i
+ \overline{\lambda^{xy}_i}\, \delta^2 \lambda^{xy}_i + 2\,\overline{\delta \lambda^{xy}_i}\,
\delta \lambda^{xy}_i \Big) \notag \\
&\quad\: -\frac{1}{4\, |\lambda^{xy}_i|^3} \: \big( \overline{\delta \lambda^{xy}_i}\, \lambda^{xy}_i
+ \overline{\lambda^{xy}_i}\, \delta \lambda^{xy}_i \big)^2 \notag \\
&= \frac{1}{|\lambda^{xy}_i|}\: \re \big( \overline{\lambda^{xy}_i}\, \delta^2 \lambda^{xy}_i \big)
+ \frac{1}{|\lambda^{xy}_i|}\: \bigg( \Big| \delta \lambda^{xy}_i \big|^2
-  \Big\{ \re  \Big(  \frac{\overline{\lambda^{xy}_i}}{|\lambda^{xy}_i|}\, \delta \lambda^{xy}_i \Big)
\Big\}^2 \bigg) \:. \label{del2abs}
\end{align}
Note that the last summand is non-negative; this corresponds to the fact that the
absolute value is a weakly convex function.

For a short compact notation, it is best to view the eigenvalues as functions of the kernel~$P(x,y)$
and its adjoint. Thus we write
\beq \label{lam1}
\lambda^{xy}_i = \lambda^{xy}_i[P(x,y)] \:.
\eeq
Moreover, we will often omit the arguments~$x$ and~$y$. Then the variations of the eigenvalues can be
written in the short form
\begin{align}
\delta \lambda_i &= D \lambda_i \big( \delta P \big) \\
\delta^2 \lambda_i &= D^2 \lambda_i \big( \delta P, \delta P \big) + D \lambda_i \big( \delta^2 P \big) \:,
\label{lam3}
\end{align}
where~$D$ and~$D^2$ denote the total derivatives (being real linear or multilinear in its arguments,
but not complex linear or multilinear).
More specifically,
\begin{align*}
D^2 \lambda_i \big( \delta P, \delta P \big) &:= \lambda^{(2)}_i + 2\, \Tr_{S_x} \big( \Lambda^{xy}_i \:\delta P(x,y)\, \delta P(y,x) \big) \\
D \lambda_i \big( \delta^2 P \big) &:= \Tr_{S_y} \big( \Lambda^{yx}_i \: P(y,x)\, \delta^2 P(x,y) \big) + \Tr_{S_x} \big( \Lambda^{xy}_i \:P(x,y)\, \delta^2 P(y,x) \big) \:.
\end{align*}
We use the same notation also for the absolute values of the eigenvalues and other composite expressions.
For example, the formula~\eqref{del2abs} can be written more compactly as
\begin{align*}
\delta^2 \big|\lambda_i \big| &= D^2 \big|\lambda_i \big|(\delta P, \delta P) +
D \big|\lambda_i \big|(\delta^2 P) \\
\intertext{with}
D^2 \big|\lambda_i \big|(\delta P, \delta P) &=
\frac{1}{|\lambda_i|}\: \re \big( \overline{\lambda_i}\, D^2 \lambda_i(\delta P, \delta P) \big) \\
&\quad\: + \frac{1}{|\lambda_i|}\: \bigg( \Big| D \lambda_i(\delta P) \big|^2
-  \Big\{ \re  \Big(  \frac{\overline{\lambda_i}}{|\lambda_i|}\, D\lambda_i(\delta P) \Big)
\Big\}^2 \bigg) \\
D \big|\lambda_i \big|(\delta^2 P) &=
\frac{1}{|\lambda_i|}\: \re \big( \overline{\lambda_i}\, D\lambda_i(\delta^2 P) \big) \:.
\end{align*}

\subsection{Second Variations of the Effective Action}
We now proceed by computing second variations of the causal action and the causal Lagrangian.
Since the considered variations~\eqref{Psivar} do not change the total volume,
the volume constraint may be disregarded. Therefore, it suffices to vary the
effective action of the form~\eqref{Seff}. A direct computation yields
\begin{align}
\delta^2 \L(x,y) &= 
\frac{1}{2n} \sum_{i,j=1}^{2n} \delta \Big( \big|\lambda^{xy}_i \big| - \big|\lambda^{xy}_j \big| \Big)\;
\delta \Big( \big|\lambda^{xy}_i \big| - \big|\lambda^{xy}_j \big| \Big) \label{t1} \\
&\quad\: +2 \kappa\: \delta \bigg( \sum_{j=1}^{2n} \big|\lambda^{xy}_j \big| \bigg)\;
 \delta \bigg( \sum_{k=1}^{2n} \big|\lambda^{xy}_k \big| \bigg) \label{t2} \\
&\quad\: + \frac{1}{2n} \sum_{i,j=1}^{2n} \Big( \big|\lambda^{xy}_i \big| - \big|\lambda^{xy}_j \big| \Big)\;
\delta^2 \Big( \big|\lambda^{xy}_i \big| - \big|\lambda^{xy}_j \big| \Big) \label{t3} \\
&\quad\: + 2 \kappa\: \bigg( \sum_{j=1}^{2n} \big|\lambda^{xy}_j \big| \bigg)\;
\delta^2 \bigg( \sum_{j=1}^{2n} \big|\lambda^{xy}_j \big| \bigg) \:. \label{t4}
\end{align}

We now want to group these terms in a way where the underlying structure becomes clear.
First, we collect all the terms involving first variations squared, i.e.\ the terms~\eqref{t1} and~\eqref{t2},
and denote them by
\begin{align}
\delta^2 \L^\lfe(x,y) &:=
\frac{1}{2n} \sum_{i,j=1}^{2n} \delta \Big( \big|\lambda^{xy}_i \big| - \big|\lambda^{xy}_j \big| \Big)\;
\delta \Big( \big|\lambda^{xy}_i \big| - \big|\lambda^{xy}_j \big| \Big) \label{lfe1} \\
&\quad\: +2 \kappa\: \delta \bigg( \sum_{j=1}^{2n} \big|\lambda^{xy}_j \big| \bigg)\;
\delta \bigg( \sum_{k=1}^{2n} \big|\lambda^{xy}_k \big| \bigg) \:. \label{lfe2}
\end{align}
This contribution is obviously non-negative.
The remaining terms~\eqref{t3} and~\eqref{t4} can be computed further with the help of~\eqref{del2abs} as well as~\eqref{dellam1} and~\eqref{dellam2}. They contain the contribution~\eqref{d2Lsep}
already considered in Section~\ref{secsep}. In order to separate this contribution, it is useful
to use the notation~\eqref{lam1}--\eqref{lam3}, making it possible to decompose the remaining terms as
\[ \eqref{t3}+\eqref{t4} = D\L(x,y) \big( \delta^2 P(x,y) \big) + R(x,y) \:, \]
where, according to~\eqref{delLdef} and similar to~\eqref{d2Lsep},
\[  D\L(x,y) \big( \delta^2 P(x,y) \big) = 2 \re \Tr_{S_xM} \!\big( Q(x,y)\, \delta^2 P(x,y)^* \big) \:, \]
whereas~$R(x,y)$ contains all the second derivatives,
\begin{align}
R(x,y) &:= \frac{1}{2n} \sum_{i,j=1}^{2n} \Big( \big|\lambda^{xy}_i \big| - \big|\lambda^{xy}_j \big| \Big)\;
D^2 \Big( \big|\lambda^{xy}_i \big| - \big|\lambda^{xy}_j \big| \Big)(\delta P, \delta P) \label{R1} \\
&\quad\: + 2 \kappa\: \bigg( \sum_{j=1}^{2n} \big|\lambda^{xy}_j \big| \bigg)\;
D^2 \bigg( \sum_{j=1}^{2n} \big|\lambda^{xy}_j \big| \bigg)(\delta P, \delta P) \:. \label{R2}
\end{align}
We thus end up with the decomposition
\beq \label{d2L}
\delta^2 \L(x,y) = \delta^2 \L^\lfe(x,y)
- 2\: \Tr_{S_x} \big( Q(x,y)\: \delta \Psi(x)^* \delta \Psi(y) \big) + R(x,y) \:.
\eeq
Using this formula for the second variations of the Lagrangian, we can compute the variation of
the causal action~\eqref{Seff}. We thus obtain the following result.

\begin{Prp} \label{prpsecond} The second variation of the effective causal action 
under variations of the wave evaluation operator of the form~\eqref{Psivar}
can be written as
\begin{align}
\delta^2 \Sact &= \int_M d\rho(x) \int_M d\rho(y)\: \delta^2 \L^\lfe(x,y) \label{D2lfe} \\
&\quad\: -4 \int_M \tr \big( (\delta \Psi)(x)^*\, \big((Q - \mathfrak{r}) (\delta \Psi) \big)(x) \big)\: d\rho(x) \label{D2Q} \\
&\quad\:  +\int_M d\rho(x) \int_M d\rho(y)\: R(x,y) \label{d2R}
\end{align}
with~$\L^\lfe(x,y)$ according to~\eqref{lfe1}, \eqref{lfe2} and~$R(x,y)$ as in~\eqref{R1}, \eqref{R2}.
\end{Prp} 
The main point of this decomposition is that~\eqref{D2lfe} and~\eqref{D2Q} are both positive,
whereas~\eqref{d2R} is small.
The positivity of~\eqref{D2lfe} is obvious from the fact that both summands~\eqref{lfe1} and~\eqref{lfe2}
are non-negative. For~\eqref{D2Q} positivity was shown in Proposition~\ref{prpQpos}.
The smallness of~\eqref{d2R} can be seen in various ways.
On a qualitative level, the smallness of the kernel~$R(x,y)$
can be seen already from the its structure as given in~\eqref{R1} and~\eqref{R2}:
The summand~\eqref{R1} is small because the term~$|\lambda^{xy}_i| - |\lambda^{xy}_j|$ vanishes
whenever~$x$ and~$y$ are spacelike separated, which includes the region where the function~$|\lambda^{xy}_i|$
is largest. The summand~\eqref{R2}, on the other hand, is small because of the smallness
of the parameter~$\kappa$.

These smallness statements can be made more quantitative by considering
specific causal fermion systems describing Minkowski space (see~\cite[Appendix~A]{jacobson} and Appendix~\ref{appMP}). In this setting, we have the following result.
\begin{Prp} \label{prp42} The second variation of the effective causal action can be written as
\begin{align}
\delta^2 \Sact &= \int_M d\rho(x) \int_M d\rho(y)\: \delta^2 \L^\lfe(x,y) \label{r0} \\
&\quad-4 \int_M \tr \big( (\delta \Psi)(x)^*\, \big((Q - \mathfrak{r}) (\delta \Psi) \big)(x) \big)\: d\rho(x)
+ (\deg \leq 2) \:. \label{r1}
\end{align}
\end{Prp}
\Proof Let us consider the terms~\eqref{R1} and~\eqref{R2} in more detail. The second variation
of the absolute values can be computed with the help of~\eqref{del2abs}.
Each of the resulting summands involves either the second variation of the eigenvalues
or the first variation of the eigenvalues squared, i.e.\
\[ \sim \delta^2 \lambda^{xy}_i \qquad \text{or} \qquad \sim \big( \delta \lambda^{xy}_i \big) \:
\big( \delta \lambda^{xy}_j \big) \]
(or their complex conjugates). The first and second variations of the eigenvalues can be computed
with the help of~\eqref{dellam1} and~\eqref{dellam2}, respectively.
The contributions involving~$\delta^2 P$ in~\eqref{dellam2} can be written as
\begin{align*}
\delta^2 \Sact &\asymp -4 \int_M \tr \big( (\delta \Psi)(x)^*\, \big((Q - \mathfrak{r}) (\delta \Psi) \big)(x) \big)\: d\rho(x) \\
&\quad\: - 4 \int_M \re \tr \big( (\delta^2 \Psi)(x)^*\, \big((Q - \mathfrak{r}) \Psi \big)(x) \big)\: d\rho(x)
\end{align*}
(here the symbol ``$\asymp$'' denotes that only the specified contribution is shown).
The last line vanishes in view of the EL equations~\eqref{ELQ}. We thus obtain the
first summand in~\eqref{r1}.

The remaining task is to show that all the remaining contributions to the second variation are of degree
at most two on the lightcone. We first note that all the remaining contributions involve two factors~$\delta P(x,y)$
or~$\delta P(y,x)$. We write this finding symbolically as follows\footnote{We note for clarity that the Lagrange multipliers are kept fixed in the variations.
This is justified by our assumption that the variation is compactly supported,
implying that the Lagrange parameters are fixed by the EL equations outside this support.},
\begin{align}
\delta^2 \Sact &= \int_M d\rho(x) \int_M d\rho(y)\: \delta^2 \L^\lfe(x,y) \label{t1n} \\
&\quad\: - 4 \int_M \tr \big( (\delta \Psi)(x)^*\, \big((Q - \mathfrak{r}) (\delta \Psi) \big)(x) \big)\: d\rho(x) \label{t2n} \\
&\quad\:+ \int_M d\rho(x) \int_M d\rho(y)\sum_{i,j=1}^{2n} \Big( \big|\lambda^{xy}_i \big| - \big|\lambda^{xy}_j \big| \Big)\:
\O \big( (\delta P)^2 \big) + \kappa\: \O \big( (\delta P)^2 \big) \:. \label{t3n} 
\end{align}
Finally, the prefactors in~\eqref{t3n} involve the scaling factors
\[ \Big( \big|\lambda^{xy}_i \big| - \big|\lambda^{xy}_j \big| \Big) \sim m^3
\qquad \text{and} \qquad \kappa \lesssim (\varepsilon m)^p + \Big( \frac{\varepsilon}{\delta} \Big)^{8-\hat{s}} \]
(for details see~\cite[eq.~(A.16)]{jacobson}) with parameters~$p>4$ and~$\hat{s} \in \{0,2\}$.
Therefore, working out these contributions in the formalism of the continuum limit, 
one finds that they can also be absorbed into the error term.
This gives the result.
\QED

\section{Approximate Decoupling of the Linearized Field Equations} \label{secdecouple}
As explained after Proposition~\ref{prpsecond}, the second variation of the causal action
involves two positive terms~\eqref{D2lfe} and~\eqref{D2Q}, but the third term~\eqref{d2R}
is not necessarily positive.
If this third term were absent, the linearized field equations would decouple, as can be
understood non-technically as follows. The solutions of the linearized field equations describe
variations which preserve the minimality of the causal action. This also means that
second variations ``in direction of the linearized field equations'' must vanish (at least if they are
well-defined, as will be analyzed in detail below).
Using positivity of both summands, we could conclude that both~\eqref{D2lfe} and~\eqref{D2Q} must vanish.
Therefore, the linearized field would satisfy separate equations, one
coming from the second variations~\eqref{D2lfe} (the so-called bosonic equations)
and one from~\eqref{D2Q} (the dynamical wave equation).

Clearly, this argument does not apply because~\eqref{d2R} is non-zero and can have an arbitrary sign.
But, by assuming that~\eqref{d2R} is very small (as has been shown in the continuum limit formalism in
Proposition~\ref{prp42} and will be analyzed and discussed in larger generality below),
we can still conclude that the linearized field equations decouple approximately.
Therefore, it makes mathematical sense to formulate two separate equations
(the bosonic equation and the dynamical wave equation), which are ``weakly coupled''
by the term~\eqref{d2R}.

In this section, we shall make this consideration mathematically precise.
One difficulty is that second variations ``in direction of the linearized field equations''
in general do not exist because the resulting spacetime integrals diverge.
This makes it necessary to consider the situation in finite time strips and consider the limiting
case that the time interval tends to infinity.
The basic construction will be carried out in Section~\ref{secapprox}.
The consequences of this approximate decoupling will be discussed in Section~\ref{secstatic}
and Appendix~\ref{appMP}.

\subsection{An Estimate in Time Strips} \label{secapprox}
We let~$\delta \Psi$ be a solution of the homogeneous linearized field equations in wave charts~\eqref{lfehom}.
By definition, this solution is defined globally in spacetime, and it does not need to decay
for large times. For this reason, $\delta \Psi$ cannot be used directly for varying the measure~$\rho$.
Instead, we need to localize in time strips, as we now explain.
First, we need to assume that we are given a continuous function~$T \in C^0(M, \R)$, referred to as the
{\em{global time function}}. The choice of the function~$T$ is arbitrary (in particular, we do not need to
specify what we mean by ``timelike'' or ``future-directed''), except for the implicit conditions which will
be posed in Definition~\ref{defboundedstrips} below.
Given a compact time interval~$I \subset \R$, the corresponding time strip~$\Omega_I$ is defined by
\beq \label{OmegaIdef}
\Omega_I :=  T^{-1} (I) \subset M \:.
\eeq

\begin{Def} \label{defboundedstrips}
The linearized solution~$\delta \Psi$ is said to be {\bf{uniformly bounded in time strips}}
if the following conditions hold:
\bitem
\item[{\rm{(i)}}] The second variation is bounded in every time strip, meaning that for every~$t_0, t_1 \in \R$,
\[ \int_{\Omega_{[t_0, t_1]}} \!\!\!\!\!\!\!\!d\rho(x) \int_{\Omega_{[t_0, t_1]}} \!\!\!\!\!\!\!\!d\rho(y)\:
\big| \delta^2 \L(x,y) \big|  < \infty \:. \]
\item[{\rm{(ii)}}] The corresponding boundary contribution defined by
\[ \int_{\Omega_{[t_0, t_1]}} \!\!\!\!\!\!\!\!d\rho(x) \int_{M \setminus \Omega_{[t_0, t_1]}} \!\!\!\!\!\!\!\!d\rho(y)\:
\big| \delta^2 \L(x,y) \big| \]
is bounded uniformly as~$t_0 \rightarrow -\infty$ and~$t_1 \rightarrow \infty$.
\eitem
\end{Def}

Under the assumptions of this definition, the following computation steps are well-defined.
We choose the time interval as~$I=[-t,t]$ with~$t>0$.
Since~$\delta \Psi$ satisfies the linearized field equations, we know that
\[ \int_{\Omega_{[-t,t]}} \!\!\!\!\!\!\!\!d\rho(x) \int_M d\rho(y)\:
\delta^2 \L(x,y) = 0 \:. \]
We now rewrite the integrals as
\begin{align*}
0 &= \int_{\Omega_{[-t,t]}} \!\!\!\!\!\!\!\!d\rho(x) \int_{\Omega_{[-t,t]}} \!\!\!\!\!\!\!\! d\rho(y)\:
\delta^2 \L(x,y)
+\int_{M \setminus \Omega_{[-t,t]}} \!\!\!\!\!\!\!\!d\rho(x) \int_{\Omega_{[-t,t]}} \!\!\!\!\!\!\!\!d\rho(y)\:
\delta^2 \L(x,y) \:.
\end{align*}
Using condition~(ii) in the above definition, we conclude that
\[ \limsup_{t \rightarrow \infty}  \int_{\Omega_{[-t,t]}} \!\!\!\!\!\!\!\!d\rho(x) \int_{\Omega_{[-t,t]}} \!\!\!\!\!\!\!\! d\rho(y)\:
\delta^2 \L(x,y) < \infty\:. \]
Now we can apply Proposition~\ref{prpsecond} and use the positivity of both~\eqref{D2lfe}
and~\eqref{D2Q} to obtain the following estimates.
\begin{Prp} \label{prppos}
Under the assumptions stated in Definition~\ref{defboundedstrips},
for second variations by the linearized solution~$\delta \Psi$,
the following inequalities hold,
\begin{align*}
\limsup_{t \rightarrow \infty} \bigg\{  \bigg| \int_{\Omega_{[-t,t]}} \!\!\!\!\!\!\!\! d\rho(x) \int_{\Omega_{[-t,t]}} \!\!\!\!\!\!\!\! d\rho(y)\: \delta^2 \L^\lfe(x,y) \bigg|
-  R(t) \bigg\} &< \infty \\
\limsup_{t \rightarrow \infty} \bigg\{ \bigg| 
\int_{\Omega_{[-t,t]}} \!\!\!\!\!\!\!\! d\rho(x) \int_{\Omega_{[-t,t]}} \!\!\!\!\!\!\!\! d\rho(y)\:
\tr \big( \delta \Psi)(x)^*\, (Q - \mathfrak{r})(x,y) \,\delta \Psi(y)\big) \bigg| - \frac{R(t)}{4} \bigg\} &< \infty \:,
\end{align*}
where
\beq \label{Rerr}
R(t) := \bigg| \int_{\Omega_{[-t,t]}} \!\!\!\!\!\!\!\! d\rho(x) \int_{\Omega_{[-t,t]}} \!\!\!\!\!\!\!\! d\rho(y)\: R(x,y) \bigg|
\eeq
and~$R(x,y)$ as in~\eqref{R1} and~\eqref{R2}.
\end{Prp}

In simple terms, these inequalities mean that contributions to
\beq \label{terms}
\delta^2 \L^\lfe(x,y) \qquad \text{and} \qquad \tr \big( \delta \Psi)(x)^*\, (Q - \mathfrak{r})(x,y) \,\delta \Psi(y)\big)
\eeq
cannot cancel each other in the linearized field equations.
Such cancellations can occur only between each contribution in~\eqref{terms} and~$R(x,y)$.
The importance of this result lies in the fact that the contribution~$R(x,y)$ is very small.
Taking this into account, the above Proposition gives us an {\em{approximate decoupling}} of the linearized field equations into a bosonic equation and the dynamical wave equation.
In view of~\eqref{lfe1} and~\eqref{lfe2}, the decoupled equations can be written as
\begin{align*}
&\int_M \frac{1}{2n} \bigg\{ \sum_{i,j=1}^{2n} \delta_x \Big( \big|\lambda^{xy}_i \big| - \big|\lambda^{xy}_j \big| \Big)\;
D \Big( \big|\lambda^{xy}_i \big| - \big|\lambda^{xy}_j \big| \Big)(\delta \Psi) \\
&\qquad\qquad +2 \kappa\: \delta_x \bigg( \sum_{j=1}^{2n} \big|\lambda^{xy}_j \big| \bigg)\;
D \bigg( \sum_{k=1}^{2n} \big|\lambda^{xy}_k \big| \bigg) (\delta \Psi) \bigg\}\: d\rho(y) = 0 \\
&\int_M (Q - \mathfrak{r})(x,y) \,\delta \Psi(y)\: d\rho(y) = 0 \:. 
\end{align*}

\subsection{Evaluation in the Static Setting} \label{secstatic}
In order to clarify what Proposition~\ref{prppos} means, we now consider the static situation.
Thus we assume that our causal fermion system as well as the linearized solution are {\em{static}}.
Here we do not need to define these notions in detail (for a formal definition 
of a static causal fermion system see~\cite[Definition~3.1]{pmt}). 
Instead, it suffices to state the precise assumptions needed for our analysis.
First, we assume that our spacetime has the product structure~$M = \R \times N$
with the first factor being the time coordinate.
Thus we represent the spacetime points as~$x = (t, \x)$ and~$y=(t', \y)$ with~$t,t' \in \R$
and~$\x, \y \in N$. Again, we identify each spacetime point with the corresponding
spacetime point operator, i.e.\ $x = (t,\x) \in \F$.
For convenience, we choose the time function~$T$ to coincide with the corresponding coordinate,
i.e.\ $T(t,\x) = t$. Next, we assume that the measure~$\rho$
is time independent in the sense that it can be written as
\[ d\rho(t,\x) = dt\, d\mu(\x) \]
with~$\mu$ a measure on~$N$. Finally, we assume that all the kernels in Proposition~\ref{prppos}
depend only on the time differences, i.e.\
\begin{gather*}
\delta^2 \L^\lfe\big((t,\x),(t',\y) \big) = A(t-t',\x, \y) \;,\qquad
R\big((t,\x),(t',\y) \big) = B(t-t',\x, \y) \\
\tr \Big( \delta \Psi)(t,\x)^*\, (Q - \mathfrak{r})\big( (t,\x),(t',\y) \big) \,\delta \Psi(t',\y) \Big)
= C(t-t',\x, \y)
\end{gather*}
with suitable functions~$A, B, C : \R \times N \times N \rightarrow \R$.
A typical example which fits into this setting is a plane wave solution in a causal fermion system
describing Minkowski space (as will be considered in more detail in Appendix~\ref{appMP}).

Under these assumptions, the leading contributions in~\eqref{Rerr} on the
left side of the inequalities in Proposition~\ref{prppos} are linear in~$t$.
For example,
\[ \int_{\Omega_{[-t,t]}} \!\!\!\!\!\!\!\! d\rho(x) \int_{\Omega_{[-t,t]}} \!\!\!\!\!\!\!\! d\rho(y)\: \delta^2 \L^\lfe(x,y)
= 2 t \int_N d\mu(\x) \int_M d\rho(t',\y)\;
\delta^2 \L^\lfe\big( (0,\x),(t,\y) \big) + \O \big( t^0 \big) \]
and similarly for the other integrals. Taking the limit~$t \rightarrow \infty$ immediately gives the following result.
\begin{Prp} In the above static setting,
\begin{align}
\bigg| \int_N d\mu(\x) \int_M d\rho(t',\y)\: \delta^2 \L^\lfe\big( (0,\x), (t',\y) \big) \bigg|
&\leq R \\
\bigg| \int_N d\mu(\x) \int_M d\rho(t',\y)\: 
\tr \big( \delta \Psi)(t,\x)^*\, (Q - \mathfrak{r})\big( (t,\x),(t',\y) \big) \,\delta \Psi(t',\y) \big) \bigg| &\leq \frac{R}{4} \label{diverge}
\end{align}
with
\beq \label{Rdef}
R := \bigg| \int_N d\mu(\x) \int_M d\rho(t',\y)\:  R\big( (0,\x), (t',\y) \big) \bigg| \:.
\eeq
\end{Prp}

In Appendix~\ref{appMP} the consequences of this statement are worked out more concretely
for the Dirac vacuum in Minkowski space.

\section{Construction of Solutions of the Dynamical Wave Equation} \label{secsoldyn}
In view of the approximate decoupling, we now focus on the dynamical wave equation.
Thus, starting from the decomposition of the second variation of the causal action in Proposition~\ref{prpsecond},
we restrict attention to the contribution~\eqref{t2} and set it equal to zero.
Likewise, setting the contribution~\eqref{t1} to zero gives the bosonic equations, which have
already been studied in detail in~\cite{nonlocal}.
Clearly, this procedure is only an approximation, because we disregard the contribution~\eqref{t3n}
which describes a coupling of the dynamical wave equation and the bosonic equations.
This procedure will be justified in Section~\ref{seccouple}, where it will be shown that
the coupling term~$R$ can indeed be treated perturbatively.

Inserting an orthonormal basis~$(e_i)$ of~$\H$, the trace in~\eqref{D2Q} can be written as
\[ \tr \big( (\delta \Psi)(x)^*\, \big((Q - \mathfrak{r}) (\delta \Psi) \big)(x) \big)
= \sum_i \Sl (\delta \psi^{e_i})(x) \:|\: \big((Q - \mathfrak{r}) (\delta \psi^{e_i}) \big)(x) \Sr_x \]
(where~$\psi^{e_i}$ is the physical wave function corresponding to~$e_i$ as defined by~\eqref{psiudef}).
In this way, the contribution~\eqref{D2Q} decomposes into a sum of non-negative terms.
Therefore, we may restrict attention to one of the summands. Thus we consider
a single wave function~$\psi \in C^0(M, SM)$ and the effective action
\beq \label{SQ}
\Sact^\text{eff} = \int_M d\rho(x) \int_M d\rho(y)\: \Sl \psi(x) \,|\, Q(x,y)\, \psi(y) \Sr 
- \mathfrak{r} \int_M \Sl \psi(x) \,|\, \psi(x) \Sr\: d\rho(x) \:.
\eeq
We now specify the regularity and decay assumptions on the kernel~$Q(x,y)$ as needed
for the subsequent analysis. 
As in Section~\ref{secapprox} we assume a global time
function~$T : M \rightarrow \R$ and consider time strips~$\Omega_I$~\eqref{OmegaIdef}.
In typical examples, the kernel~$Q(x,y)$ is smooth (in view of the ultraviolet regularization)
and decays if~$x$ and~$y$ are far apart. Therefore, it is sensible to make the following assumptions.

Similar to the construction of the Krein space structures in~\cite[\S1.1.5]{cfs}, on the spin spaces we
introduce the scalar product
\beq \label{spinscalar}
\lla .|. \rra_x \::\: S_xM \times S_xM \rightarrow \C\:,\qquad 
\lla \psi | \phi \rra_x := \la \psi \,|\, |x|\, \phi \ra_\H
\eeq
(where we make use of the fact that~$S_xM \subset \H$). The corresponding norm is denoted
by~$\|.\|_x$. These norms also induce a corresponding sup-norm on the linear operators~$\Lin(S_yM, S_xM)$
mapping from the spin space~$S_yM$ to~$S_xM$; for notational simplicity we denote it by
\[ \|A \| := \sup_{\phi \in S_yM \text{ with } \|\phi\|_y=1} \| A \phi \|_x \:,\qquad
A \in \Lin(S_yM, S_xM)\:. \]
\begin{Def} \label{defunibound}
We assume that the kernel~$Q(x,y)$ is {\bf{uniformly bounded in $L^1$}} in the sense that there is a
constant~$c>0$ such that
\[ \int_M \|Q(x,y)\|\: d\rho(y) < c \qquad \text{for all~$x \in M$}\:. \]
\end{Def}

\begin{Def} \label{defrange}
The kernel~$Q(x,y)$ has {\bf{finite time range}}
if there is a parameter~$\delta>0$ such that
\[ Q\big( (t,\x),(t',\y) \big) = 0 \qquad \text{if~$|t-t'| > \delta$}\:. \]
\end{Def}
This assumption can be understood in physical terms from the fact that the kernel~$Q(x,y)$
decays on the Compton scale
(again for Dirac systems in Minkowski space as constructed in~\cite[Section~1.2]{cfs} or~\cite[Section~5.5]{intro});
see also~\cite[Section~9.1]{intro}.
Therefore, the assumption of finite time range is satisfied approximately,
with an error which can be made arbitrarily small by choosing~$\delta$ sufficiently large.
This decay property has been used in various mathematical formulations in the literature
(like {\em{compact range}} in~\cite{noncompact, intro} or 
{\em{decays on the scale~$\delta$}} in~\cite{matter}).
For our purposes, the notion of finite time range seems most suitable.

\subsection{Construction of Inhomogeneous Solutions} \label{secinhom}
We now give a method for constructing solutions of the dynamical wave equation.
This method was proposed abstractly for the linearized field equations
in~\cite[Section~5]{nonlocal}. We now work it out in more detail.
We point out that the following constructions do not rely on energy methods
(as used previously for the dynamical wave equation in~\cite{dirac}).
Instead, the method given here works exclusively with the positivity of second variations.

We choose a time interval~$I=[t_0, t_1]$ and let~$\Omega := \Omega_I$ be the corresponding time strip.
We restrict our effective action to this time strip,
\begin{align*}
\Sact^\text{eff}_\Omega &: C^0_0(\Omega, S\Omega) \times C^0_0(\Omega, S\Omega) \rightarrow \C \:, \\
\Sact^\text{eff}_\Omega(\psi, \phi) &:= \int_\Omega d\rho(x) \int_\Omega d\rho(y)\: \Sl \psi(x) \,|\, Q(x,y)\, \phi(y) \Sr
- \mathfrak{r} \int_\Omega \Sl \psi(x) \,|\, \psi(x) \Sr\: d\rho(x)  \:.
\end{align*}
Thus~$\Sact^\text{\eff}_\Omega$ is a sesquilinear form on~$C^0(\Omega, S\Omega)$.
It is positive semi-definite according to Proposition~\ref{prpQpos}.
We point out for clarity that the time strip will be fixed throughout our construction.
This is an important point, because we do not expect
our method to be uniform in the size of~$I$.

On~$C^0_0(\Omega, S\Omega)$ we introduce the $L^2$-scalar product
\beq \label{HOdef}
\la \psi | \phi \ra_\Omega := \int_\Omega \lla \psi(x) \,|\, \phi(x) \rra_x \: d\rho(x) \:,
\eeq
where in the integrand we work with the scalar product~\eqref{spinscalar} on the spin spaces.
Taking the completion, we obtain a Hilbert space denoted by~$(\H_\Omega, \la .|. \ra_\Omega)$.
The norm on this Hilbert space is denoted by~$\|.\|_\Omega$.
\begin{Lemma}
The effective action~$\Sact^\text{eff}_\Omega$, \eqref{Seff}, is bounded with respect to the scalar product~\eqref{HOdef}, i.e.\ there is a constant~$c>0$ with
\[ \big| \Sact^\text{eff}_\Omega(\psi, \phi) \big| \leq c\: \| \psi \|_\Omega\:  \| \psi \|_\Omega \qquad
\text{for all~$\psi, \phi \in \H_\Omega$}\:. \]
\end{Lemma}
\Proof Using the Schwarz inequality on~$L^2(M, SM)$, we obtain
\begin{align*}
&\big| \delta^2 \Sact^\text{eff}_\Omega(\psi, \phi) \big|
\leq \int_M d\rho(x) \int_M d\rho(y) \: \|\psi(x)\|_x\: \|Q(x,y)\|\: \|\phi(y)\|_y \\
&= \int_M d\rho(x) \int_M d\rho(y) \: \Big( \|\psi(x)\|_x\: \sqrt{\|Q(x,y)\|} \Big)\: \Big( \sqrt{\|Q(x,y)\|}\: \|\phi(y)\|_y \Big) \\
&\leq \int_M d\rho(x) \int_M d\rho(y) \: \sup_{\tilde{y} \in M} \Big( \|\psi(x)\|_x\: \sqrt{\|Q(x,\tilde{y})\|} \Big)\: 
\sup_{\tilde{x} \in M} \Big( \sqrt{\|Q(\tilde{x},y)\|}\: \|\phi(y)\|_y \Big) \\
&\leq \|\psi\|_{L^2(\Omega)}\: \sup_{\tilde{y} \in M} \sqrt{ \|Q(.,\tilde{y}) \|_{L^2(\Omega)}}\;
\|\phi\|_{L^2(\Omega)}\: \sup_{\tilde{x} \in M} \sqrt{\|Q(\tilde{x},.) \|_{L^2(\Omega)}} \\
&\leq c\: \|\psi\|_{L^2(\Omega)}\: \|\phi\|_{L^2(\Omega)} \:,
\end{align*}
where~$c$ is the constant in Definition~\ref{defunibound}.
\QED
Using this lemma, the second variations can be represented relative to the $L^2$-scalar product by
\[ \Sact^\text{eff}_\Omega(\phi, \psi) = \la \phi \,|\, \mathcal{Q}\: \psi \ra_\Omega \qquad
\text{for all~$\phi, \psi \in \H_\Omega$}\:, \]
with a linear operator~$\mathcal{Q}$ which is bounded and symmetric
(note that, for convenience, the Lagrange multiplier term~$\mathfrak{r}$ was absorbed into~$\mathcal{Q}$).

In this setting, one can construct solutions of the inhomogeneous equations abstractly as follows.
Let~$\phi \in \H_\Omega$. Assuming that~$\phi$ lies in the image of the operator~$\mathcal{Q}$,
there is a vector~$\psi \in \H_\Omega$ with
\beq \label{Qeq}
\mathcal{Q} \psi = \phi \:.
\eeq
This vector is a solution of the dynamical wave equation.
Clearly, this procedure raises the questions under which conditions~$\phi$ lies in the image of~$\mathcal{Q}$
and whether the solution is unique. These questions can be answered as follows.
Since~$\Sact^\text{eff}_\Omega$ is positive semi-definite, we know that~$\mathcal{Q} \geq 0$.
But this operator could have a kernel. In this case, the solution is unique only up to vectors in this kernel.
If the operator~$\mathcal{Q}$ is strictly positive (i.e.\ $\mathcal{Q} \geq \nu$ with~$\nu>0$), then it is surjective,
giving existence of a unique solution for every~$\phi \in \H_\Omega$.
We expect that, in typical applications, strict positivity should hold in every time strip.
But there is no general proof of this conjecture.

For the following construction, it suffices to choose let~$\psi$ be any solution of the equation~\eqref{Qeq}.
As we shall see, different solutions will differ only by vectors in the kernel of the inner product~$\la .|. \ra^t$,
implying that they will give rise to the same vector in the extended Hilbert space
(see Lemma~\ref{lemmakernel}).
In order to ensure existence of solutions, we must be careful to only consider inhomogeneities
of the following form.
\begin{Def} \label{defadmissible}
An inhomogeneity~$\phi \in \H_\Omega$ is called {\bf{admissible}} if it lies in the image of the
operator~$\mathcal{Q}$.
\end{Def}

\subsection{Construction of Homogeneous Solutions} \label{sechom}
\begin{figure}
\psset{xunit=.5pt,yunit=.5pt,runit=.5pt}
\begin{pspicture}(406.52496578,285.69136011)
{
\newrgbcolor{curcolor}{0.80000001 0.80000001 0.80000001}
\pscustom[linestyle=none,fillstyle=solid,fillcolor=curcolor]
{
\newpath
\moveto(1.2638643,178.49329505)
\lineto(404.63206542,178.7828917)
\lineto(404.63210325,107.46144579)
\lineto(1.75823123,107.22603552)
\closepath
}
}
{
\newrgbcolor{curcolor}{0.80000001 0.80000001 0.80000001}
\pscustom[linewidth=1.61127723,linecolor=curcolor]
{
\newpath
\moveto(1.2638643,178.49329505)
\lineto(404.63206542,178.7828917)
\lineto(404.63210325,107.46144579)
\lineto(1.75823123,107.22603552)
\closepath
}
}
{
\newrgbcolor{curcolor}{0 0 0}
\pscustom[linewidth=2.75905519,linecolor=curcolor]
{
\newpath
\moveto(25.69985008,0.01140043)
\lineto(23.48501669,268.30990625)
}
}
{
\newrgbcolor{curcolor}{0 0 0}
\pscustom[linestyle=none,fillstyle=solid,fillcolor=curcolor]
{
\newpath
\moveto(15.79181111,264.38358913)
\lineto(23.34153089,285.69136172)
\lineto(31.24199374,264.51113207)
\curveto(26.71024232,268.33653084)(20.2791038,268.28344109)(15.79181111,264.38358913)
\closepath
}
}
{
\newrgbcolor{curcolor}{0 0 0}
\pscustom[linewidth=2.0031495,linecolor=curcolor]
{
\newpath
\moveto(2.27506772,16.00278563)
\lineto(404.75289827,16.82754657)
}
}
{
\newrgbcolor{curcolor}{0 0 0}
\pscustom[linewidth=2.0031495,linecolor=curcolor]
{
\newpath
\moveto(0.00205606,251.24245177)
\lineto(402.4798526,252.06721271)
}
}
{
\newrgbcolor{curcolor}{0.65098041 0.65098041 0.65098041}
\pscustom[linestyle=none,fillstyle=solid,fillcolor=curcolor]
{
\newpath
\moveto(113.07572031,88.35491287)
\curveto(74.77528063,80.28979035)(40.58248063,70.2289261)(56.77053354,57.431442)
\curveto(72.95858646,44.63395791)(139.52488063,29.10118814)(207.32464252,29.48197555)
\curveto(275.12440441,29.86280074)(344.14951559,46.1570924)(349.16559118,61.98921964)
\curveto(354.18166677,77.82134688)(295.18800756,93.18940862)(244.48533921,97.83924799)
\curveto(193.78267087,102.48908736)(151.37616,96.42003539)(113.07572031,88.35491287)
\closepath
}
}
{
\newrgbcolor{curcolor}{0 0 0}
\pscustom[linewidth=0.99999871,linecolor=curcolor]
{
\newpath
\moveto(113.07572031,88.35491287)
\curveto(74.77528063,80.28979035)(40.58248063,70.2289261)(56.77053354,57.431442)
\curveto(72.95858646,44.63395791)(139.52488063,29.10118814)(207.32464252,29.48197555)
\curveto(275.12440441,29.86280074)(344.14951559,46.1570924)(349.16559118,61.98921964)
\curveto(354.18166677,77.82134688)(295.18800756,93.18940862)(244.48533921,97.83924799)
\curveto(193.78267087,102.48908736)(151.37616,96.42003539)(113.07572031,88.35491287)
\closepath
}
}
{
\newrgbcolor{curcolor}{0.65098041 0.65098041 0.65098041}
\pscustom[linestyle=none,fillstyle=solid,fillcolor=curcolor]
{
\newpath
\moveto(124.53023244,233.54576263)
\curveto(98.76555969,226.82058468)(79.12572472,217.65683287)(97.09657323,207.08784137)
\curveto(115.06741795,196.51884988)(170.64641008,184.5456883)(219.02568189,188.22671003)
\curveto(267.4049537,191.90771665)(308.57874142,211.24243791)(306.55276346,223.95468232)
\curveto(304.52678929,236.66692673)(259.30129134,242.75516736)(220.74232819,243.65621807)
\curveto(182.18336882,244.55726877)(150.2949052,240.27094059)(124.53023244,233.54576263)
\closepath
}
}
{
\newrgbcolor{curcolor}{0 0 0}
\pscustom[linewidth=0.99999871,linecolor=curcolor]
{
\newpath
\moveto(124.53023244,233.54576263)
\curveto(98.76555969,226.82058468)(79.12572472,217.65683287)(97.09657323,207.08784137)
\curveto(115.06741795,196.51884988)(170.64641008,184.5456883)(219.02568189,188.22671003)
\curveto(267.4049537,191.90771665)(308.57874142,211.24243791)(306.55276346,223.95468232)
\curveto(304.52678929,236.66692673)(259.30129134,242.75516736)(220.74232819,243.65621807)
\curveto(182.18336882,244.55726877)(150.2949052,240.27094059)(124.53023244,233.54576263)
\closepath
}
}
{
\newrgbcolor{curcolor}{0 0 0}
\pscustom[linewidth=1.13385831,linecolor=curcolor]
{
\newpath
\moveto(1.7068422,179.0186357)
\lineto(404.1846463,179.84339665)
}
}
{
\newrgbcolor{curcolor}{0 0 0}
\pscustom[linewidth=1.13385831,linecolor=curcolor]
{
\newpath
\moveto(2.9575748,106.16566814)
\lineto(405.43540535,106.99042909)
}
}
{
\newrgbcolor{curcolor}{0 0 0}
\pscustom[linewidth=1.13385831,linecolor=curcolor,linestyle=dashed,dash=0.89999998 0.89999998]
{
\newpath
\moveto(1.20477732,233.87151255)
\lineto(403.68257386,234.6962735)
}
}
{
\newrgbcolor{curcolor}{0 0 0}
\pscustom[linewidth=1.13385831,linecolor=curcolor,linestyle=dashed,dash=0.60000002 0.60000002]
{
\newpath
\moveto(1.82106709,37.67471003)
\lineto(404.29890142,38.49947098)
}
}
{
\newrgbcolor{curcolor}{0 0 0}
\pscustom[linewidth=1.13385831,linecolor=curcolor,linestyle=dashed,dash=0.89999998 0.89999998]
{
\newpath
\moveto(4.04604472,199.20804043)
\lineto(406.5238337,200.03280137)
}
}
{
\newrgbcolor{curcolor}{0 0 0}
\pscustom[linewidth=1.13385831,linecolor=curcolor,linestyle=dashed,dash=0.60000002 0.60000002]
{
\newpath
\moveto(2.9575748,85.97627098)
\lineto(405.43540535,86.80103192)
}
\rput[bl](415, 245){$t_1$}
\rput[bl](415, 225){$t_1-\delta$}
\rput[bl](415, 170){$t_{\max}$}
\rput[bl](415, 192){$t_{\max}+\delta$}
\rput[bl](415, 100){$t_{\min}$}
\rput[bl](415, 76){$t_{\min}-\delta$}
\rput[bl](415, 8){$t_0$}
\rput[bl](415, 30){$t_0+\delta$}
\rput[bl](180, 53){$\supp \phi_0$}
\rput[bl](180, 208){$\supp \phi_1$}
}
\end{pspicture}
\caption{Construction of homogeneous solutions in a time strip.}
\label{figstrip}
\end{figure}
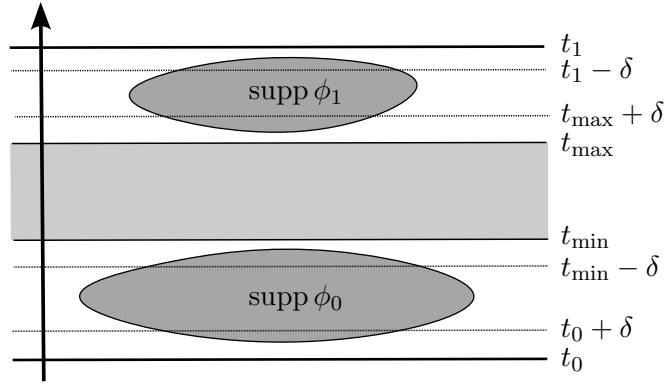

Our method for constructing homogeneous solutions is to localize the inhomogeneity
in a small time strip in the future. Thus we choose a time~$t_{\max} \in [t_0, t_1]$ close to the
final time~$t_1$ (see Figure~\ref{figstrip}) and consider an admissible function~$\phi \in \H_\Omega$
which is supported in the time strip between~$t_{\max}$ and~$t_1$, i.e.\
\beq \label{inhomfuture}
\supp \phi \subset \Omega_{[t_{\max}, t_1]} \:.
\eeq
Then the resulting solution~$\psi$ of the inhomogeneous equation~\eqref{Qeq}
is homogeneous up to time~$t_{\max}$, i.e.\
\beq \label{Qhom}
(\mathcal{Q} \psi)(t,\x)=0 \qquad \text{for all~$t \in [t_0, t_{\max}]$}\:.
\eeq

\subsection{Arranging Positivity of the Commutator Inner Product} \label{secpositive}
We now consider the solution~$\psi \in \H_\Omega$ in~\eqref{Qhom} and extend it by zero to all of~$M$.
Then~$\psi$ satisfies the dynamical wave equation
\beq \label{dyn}
Q \psi = \mathfrak{r}\, \psi \qquad \text{in~$\Omega$}\:.
\eeq
We now compute the corresponding commutator inner product~\eqref{cip} for~$\Omega$ as the
past at time~$t$, i.e.\
\beq \label{spt}
\la .|. \ra^t := \la .|. \ra^{\Omega_{(-\infty, t]}} \:.
\eeq
A direct computation using~\eqref{dyn} yields that the commutator inner product is independent of~$t
\in [t_0, t_{\max}]$.
Noting that it vanishes trivially at time~$t_0$ (simply because in~\eqref{cip} both integrals 
over the past of~$t_0$ are zero), we conclude that
\beq \label{cipzero}
\la \psi | \psi \ra^t = 0 \qquad \text{for all~$t \in [t_0, t_{\max}]$}\:.
\eeq
Therefore, all the homogeneous solutions constructed so far suffer from the shortcoming that
their commutator inner product is trivial. In order to cure this problem, we now proceed by
modifying the solutions with an additional inhomogeneity near the initial time~$t_0$.

In order to improve the situation, we now consider more general homogeneous solutions.
To this end, we choose the inhomogeneity~$\phi$ in~\eqref{Qeq} as
\beq \label{phi01def}
\begin{split}
&\phi = \lambda \phi_0 + \phi_1 \qquad \text{with} \qquad \text{$\phi_{0\!/\!1} \in \H_\Omega$ admissible and} \\
&\supp \phi_0 \subset \Omega_{[t_0, t_{\min}]}, \;
\supp \phi_1 \subset \Omega_{[t_{\max}, t_1]} \:,
\end{split}
\eeq
where~$t_{\min}$ and~$t_{\max}$ are chosen as in Figure~\ref{figstrip}.
The resulting inhomogeneous solution~$\psi$ of~\eqref{Qeq} is homogeneous away from the boundaries,
\beq \label{Qhom2}
(\mathcal{Q} \psi)(t,\x)=0 \qquad \text{for all~$t \in [t_{\min}, t_{\max}]$}\:.
\eeq
If~$\lambda=0$, we are back in the setting of the previous
section, and the commutator inner product is trivial. However, linearly in~$\lambda$, the situation changes,
because for any~$t \in [t_{\min}+\delta, t_{\max}-\delta]$,
\begin{align*}
\la \psi | \psi \ra^t &= \la \psi | \psi \ra^t - \la \psi | \psi \ra^{t_0}
= \int_{t_0}^t\: \frac{d}{d\tau} \la \psi | \psi \ra^\tau \:d\tau \\
&= -2i \int_{\Omega_{[t_0, t]}} \!\!\!\!\!\! \Sl \psi(x) \,|\, (Q \psi)(x) \Sr_x\: d\rho(x)
+ 2i \int_{\Omega_{[t_0, t]}} \!\!\!\!\!\! \Sl (Q \psi)(y) \,|\, \psi(y) \Sr_y\: d\rho(y) \\
&= 4 \im \int_{\Omega_{[t_0, t]}} \!\!\!\!\!\! \Sl \psi(x) \,|\, \phi_0(x) \Sr_x\: d\rho(x) \:.
\end{align*}
Now the commutator inner product is in general non-zero.
With the following construction we can even arrange it to be positive:
We first choose~$\phi_0=0$ and denote the corresponding solution by~$\psi^{(1)}$.
Now we choose
\beq \label{phi0lin}
\phi_0(x) = \lambda\,\frac{i}{4}\: s_x\, \psi^{(0)}(x) \: \chi_{\Omega_{[t_0, t_{\min}]}}
\eeq
(with~$s_x$ the Euclidean sign operator). Here~$\lambda$ is a small parameter. 
Then, linearly in~$\lambda$,
\begin{align}
\la \psi | \psi \ra^t &=
4 \lambda \im \int_{\Omega_{[t_0, t_{\min}]}} \!\!\!\!\!\! \Sl \psi^{(0)}(x) \,|\, \frac{i}{4}\:s_x\, \psi^{(0)}(x) \Sr_x\: d\rho(x) 
+ \O \big(\lambda^2 \big) \notag \\
&= \lambda \int_{\Omega_{[t_0, t_{\min}]}} \!\!\!\!\!\! \lla \psi^{(0)}(x) \,|\, \psi^{(0)}(x) \rra_x\: d\rho(x) 
+ \O \big(\lambda^2 \big) \notag \\
&= \lambda \:\la \psi^{(0)} \,|\, \psi^{(0)} \ra_{\Omega_{[t_0, t_{\min}]}} + \O \big(\lambda^2 \big) \:. \label{sprodlin}
\end{align}

Before going on, we comment on the physical meaning of this construction.
\begin{Remark} (possible physical interpretation) \label{rembigbang} {\em{
We now try to extrapolate from the time strip shown in Figure~\ref{figstrip} to our Universe.
Although being somewhat speculative, the following consideration seems helpful for the understanding.
It will not be used later in this paper.

In Section~\ref{sechom} we constructed solutions of the dynamical wave equation
with an inhomogeneity supported in the future of the time strip of interest~\eqref{inhomfuture}.
The purpose of the inhomogeneity is to obtain a non-trivial solution. Therefore, $\phi_0$
can be understood as parametrizing the corresponding solution~$\psi$.
On all the solutions obtained in this way, the commutator inner product vanishes~\eqref{cipzero}.
Thinking of~$t_0$ as the time of the big bang, this means that for all solutions in spacetime
which are homogeneous up to the big bang, the commutator inner product is zero.
In order to obtain a non-trivial commutator inner product, we had to introduce an inhomogeneity
also near~$t_0$ (see~\eqref{phi01def}).
Again thinking of~$t_0$ as the time of the big bang, this means that the spacetime singularity
at the big bang must acts as an effective inhomogeneity.
In other words, the dynamical wave equation should not hold all the way up to and including the
big bang. Instead, the big bang singularity should give rise to an effective inhomogeneity.
Taking this picture seriously, the positivity of the commutator inner product is a remnant of the big bang.
Ultimately, the positivity of the commutator inner product should be explained from the
structure of the big bang singularity.
}} \QEDrem
\end{Remark}

\section{Construction of the Extended Hilbert Space} \label{secexhil}
Following Definition~\ref{defSLrep} we let~$\H^\fermi \subset \H$ be a subspace of the Hilbert
space~$(\H, \la .|. \ra_\H)$ on which the commutator inner product represents the scalar product.
Our goal is to extend this Hilbert space to the {\em{extended Hilbert space}} denoted
by~$(\H^t_\rho, \la .|. \ra^t_\rho)$. By rescaling the scalar product~$\la .|. \ra_\H$, we can
arrange that the constant~$c$ in~\eqref{Ccond} is equal to one.
In order to construct the wave functions in the extended Hilbert
space, we want to employ the method developed in Section~\ref{secsoldyn}
of considering inhomogeneous solutions in a time strip whose inhomogeneities are supported in small
strips near the initial and final time (as depicted in Figure~\ref{figstrip}).

The main remaining task is to generalize the above methods in such a way that every vector in~$\H^\fermi$
is associated to a vector in~$\H^t_\rho$, giving rise to an isometric embedding
\[ 
\iota : \H^\fermi \hookrightarrow \H^t_\rho \:. \]
In preparation, we generalize the construction of the previous section. To this end, given
admissible $\phi_0, \phi_1 \in \H_\Omega$ supported near the past respectively future boundary, i.e.\
\[ \supp \phi_0 \subset \Omega_{[t_0, t_{\min}]} \:,\qquad \supp \phi_1 \subset \Omega_{[t_{\max}, t_1]} \:, \]
we consider the corresponding inhomogeneous solutions~$\phi_0$ and~$\phi_1$ defined by
\beq \label{Q01}
{\mathcal{Q}} \,\psi_{0\!/\!1} = \phi_{0\!/\!1} \:.
\eeq
In this way, we have constructed two homogeneous solutions in the intermediate time strip, i.e.\
\[  \mathcal{Q} \,\psi_{0\!/\!1} = 0 \qquad \text{in~$\Omega_{(t_{\min}, t_{\max})}$}\:. \]
Clearly, these solutions are defined only up to vectors in the kernel of~${\mathcal{Q}}$.
Before studying this non-uniqueness issue, we note that the argument leading to~\eqref{cipzero} showed that the commutator inner product~$\la \psi_1 | \psi_1 \ra^t$ vanishes for all~$t \in (t_{\min}, t_{\max})$.
Reverting the time direction, one sees with the same argument that also~$\la \psi_0 | \psi_0 \ra^t = 0$.
However, the inner product~$\la \psi_0 | \psi_1 \ra^t$ is in general non-zero.
Therefore, denoting all the solutions obtained with the above construction by
\[ \K_0 := \{ \text{$\psi_0$ satisfying~\eqref{Q01}} \} \:, \qquad \K_1 := \{ \text{$\psi_1$ satisfying~\eqref{Q01}} \} \:, \]
and taking their direct sum,
\[ \K := \K_0 \oplus \K_1 \;\subset\; \H_\Omega\:, \]
we obtain a complex vector space~$\K$ endowed with a sesquilinear form~$\la .|. \ra^t$,
spanned by two neutral subspaces~$\K_0$ and~$\K_1$.
We also note that~$\K$ carries the topology
inherited from the Hilbert space~$\H_\Omega$ (introduced after~\eqref{HOdef}).
It is an important observation that vectors in the kernel of~$\mathcal{Q}$ even lie in the kernel of
this sesquilinear form:
\begin{Lemma} \label{lemmakernel}
Suppose that the operator~$\mathcal{Q}$ has a non-trivial kernel, i.e.\ that there is
a nonzero vector~$\varphi \in \H_\Omega$ with~$\mathcal{Q}\varphi=0$. Then
\[ \la \varphi | \chi \ra^t = 0 \qquad \text{for all~$\chi \in \K$ and~$t \in (t_{\min}, t_{\max})$}\:. \]
\end{Lemma}
\Proof It suffices to consider the cases~$\chi=\psi_0$ and~$\chi=\psi_1$.
We begin with the case~$\chi=\psi_1$.
Since the corresponding inhomogeneity~$\phi_1$ is supported near the
future boundary, we know that
\[ \la \varphi | \psi_1 \ra^t = \la \varphi | \psi_1 \ra^{t_0} = 0 \:. \]
The case~$\chi=\psi_0$ follows in the same way after reverting the time direction. 
This concludes the proof.
\QED

Our next goal is to embed~$\H^\fermi$ into~$\K$. To this end, given a physical wave function we need to construct
corresponding inhomogeneities~$\phi_0$ and~$\phi_1$.
Here is where the assumption of finite time range will be needed (Definition~\ref{defrange}).
We choose the time strips near the past and future sufficiently large compared to the time range~$\delta$;
more precisely, we need to arrange that
\[ t_1-t_{\max},\; t_{\min} - t_0 > 2 \delta \]
(see again Figure~\ref{figstrip}).
Now we choose two cutoff functions, one in the past and one in the future,
\beq \label{eta01def}
\begin{split}
\eta_0 \in C^0([t_0, t_1], \R) \quad &\text{with} \quad
0 \leq \eta_0 \leq1\:, \quad\supp \eta_0 \subset (t_0+\delta, t_1]\:,\quad \eta_0|_{(t_{\min}-\delta, t_1]} \equiv 1 \\
\eta_1 \in C^0([t_0, t_1], \R) \quad &\text{with} \quad
0 \leq \eta_1 \leq1\:, \quad\supp \eta_1 \subset [t_0, t_1-\delta)\:,\quad \eta_0|_{[t_0, t_{\max}+\delta]} \equiv 1\:.
\end{split}
\eeq
We introduce corresponding cutoff functions in spacetime by setting~$\eta_{0\!/\!1}(t,\x) := \eta_{0\!/\!1}(t)$.
Let~$u \in \H^\fermi$ and~$\psi^u$ the corresponding physical wave function. Then
\beq \label{phi01concrete}
\left.\begin{matrix}
\phi_0 := \mathcal{Q} \big( \eta_0\, \psi^u \big) \\[0.2em]
\phi_1 := \mathcal{Q} \big( \eta_1\, \psi^u \big) \end{matrix} \;\;\right\}
\quad \text{is supported in} \quad \left\{ \;\;\begin{matrix} \!\!\!\Omega_{[t_0, t_{\min}]} \\[0.2em]
\Omega_{[t_{\max}, t_1]} \:. \end{matrix} \right.
\eeq
By construction, we know that the vector~$\phi_0+\phi_1 \in \H_\Omega$ 
(again with~$\Omega := \Omega_{[t_0, t_1]}$)
is admissible (as defined in Definition~\ref{defadmissible}),
simply because~$\mathcal{Q} \big( \eta_0\, \eta_1 \psi^u \big) = \phi_0+\phi_1$).
In this way, every physical wave function in~$\H^\fermi$ can be obtained from a corresponding
inhomogeneous solution.

One should keep in mind that the last argument only yields that~$\phi_0+\phi_1$ is admissible,
but it is not guaranteed that~$\phi_0$ and~$\phi_1$ are separately admissible.
Therefore, we need to take this as an additional
assumption which, for technical simplicity, we formulate as follows.
\begin{Def} \label{defHadmissible}
The subspace~$\H^\fermi \subset \H$ is said to be {\bf{admissible}} if for every~$u \in \H^\fermi$,
the corresponding inhomogeneities~$\phi_0$ and~$\phi_1$ (as defined by~\eqref{phi01concrete})
are both admissible (in the sense of Definition~\ref{defadmissible}).
\end{Def} \noindent
This technical assumption seems unproblematic because it
can be arranged by making~$\H^\fermi$ smaller if necessary.
Denoting the corresponding inhomogeneous solutions by~$\psi^u_0$ and~$\psi^u_1$,
we can decompose every physical wave function~$\psi^u$ with~$u \in \H^\fermi$ uniquely as
\[ \psi^u = \psi^u_0 + \psi^u_1 \qquad \text{with} \qquad \psi^u_0 \in \K_0,\; \psi^u_1 \in \K_1 \:. \]
This gives rise to a natural isometric embedding
\[ \iota : \H^\fermi \hookrightarrow \K \:, \qquad u \mapsto \psi^u_0 + \psi^u_1 \:. \]

The remaining task is to choose a $\H^t_\rho$ as a maximal positive subspace of~$\K$
containing~$\iota(\H^\fermi)$. To this end, we first consider the orthogonal complement
\[ \iota(\H^\fermi)^\perp \subset \K \]
(note that, since~$\H^\fermi$ is a Hilbert space, we may form the orthogonal projection onto it).
In this space we choose a maximal positive subspace as follows. We begin with the neutral subspace
\beq \label{K1ext}
\K_1 \cap \iota(\H^\fermi)^\perp \:.
\eeq
A wave function in this vector space, denoted by~$\psi^{(0)}$, can be represented by construction
as~$\psi^{(0)} = \mathcal{Q}^{-1}(\phi_1)$, where~$\phi_1$ is supported in the
time strip~$\Omega_{(t_{\max}, t_1)}$ near the future boundary.
Following the construction in Section~\ref{secpositive}, we now perturb this solution
linearly with an inhomogeneity~$\phi_0$ supported near the past boundary
(see~\eqref{phi01def} and~\eqref{phi0lin}).
According to~\eqref{sprodlin}, the inner product~$\la .|. \ra^t$ is positive definite on the 
resulting solution~$\psi := \psi^{(0)} + \lambda \psi^{(1)}$.
Moreover, the vector~$\psi$ is orthogonal to~$\H^\fermi$ up to terms linear in~$\lambda$.
This implies that the inner product~$\la .|. \ra^t$ is also positive on~$\H^\fermi \oplus \text{span}(\psi)$
for sufficiently small~$\lambda$.

In order to truncate the higher orders in a consistent way, it is convenient to
restrict attention to the resulting pair~$(\psi^{(0)}, \psi^{(1)}) \in \H_\Omega \times \H_\Omega$.
Both~$\psi^{(0)}$ and~$\psi^{(1)}$ are solutions of the homogeneous dynamical wave equation away from
the inhomogeneities, i.e.\
\[ (Q - \mathfrak{r} )\, \psi^{(0)} = 0 = (Q - \mathfrak{r} )\, \psi^{(1)}
\qquad \text{in~$\Omega_{(t_{\min}+\delta, t_{\max}-\delta)}$}\:. \]
We define the scalar product on these pairs as the linear contribution to the
commutator inner product,
\[ \big\la (\psi^{(0)}, \psi^{(1)}) \,\big|\, (\varphi^{(0)}, \varphi^{(1)}) \big\ra^t
:= \lambda \:\Big( \la \psi^{(0)} | \varphi^{(1)} \ra^t + \la \psi^{(1)} | \varphi^{(0)} \ra^t \Big) \]
(with the sesquilinear forms on the right side given by~\eqref{spt} and~\eqref{cip}).
This sesquilinear form is positive semi-definite in view of~\eqref{sprodlin}.
We thus obtain the following result.
\begin{Thm} \label{thmmain} Assume that the kernel~$Q(x,y)$ is uniformly bounded in~$L^1$
(see Definition~\ref{defunibound}) and that there is a global time function~$T \in C^0(M, \R)$
such that the kernel~$Q(x,y)$ has finite time range (see Definition~\ref{defrange}).
Moreover, given a time strip~$[t_{\min}, t_{\max}]$, we assume that there are cutoff
functions~$\eta_0$ and~$\eta_1$ as in~\eqref{eta01def} such
that~$\H^\fermi$ is admissible (see Definition~\ref{defHadmissible}). Then the
{\bf{extended Hilbert space}}~$\H^t_\rho$ can be defined for any~$t \in (t_{\min}, t_{\max})$ as the completion
\[ \H^t_\rho := \overline{\big\{ (\psi^{(0)}, \psi^{(1)}) \text{ as above} \big\}
\oplus \iota \big(\H^\fermi \big) }^{\la .|. \ra^t} \:. \]
The time evolution can be described by a unitary operator
\[ U^{t'}_t \::\: \H^t_\rho \rightarrow \H^{t'}_\rho \quad \text{unitary} \:. \]
\end{Thm}
\Proof It remains to show that the time evolution can be described by a unitary operator.
This follows immediately from the fact that, 
by construction, the commutator inner product is time independent.
\QED 
Identifying the Hilbert spaces under this unitary mapping, in what follows we can omit the index~$t$ and
denote the extended Hilbert space simply by~$(\H_\rho, \la .|. \ra)$.

This construction requires a brief explanation. We first note that, according to Lemma~\ref{lemmakernel},
the extended Hilbert space remains unchanged if its vectors are modified by vectors in the kernel of
the operator~$\mathcal{Q}$. We also point out that all the constructions up to~\eqref{K1ext} are canonical.
However, the choice of the linear perturbation~\eqref{phi0lin} is {\em{not}} canonical,
because it depends on the choice
of the inhomogeneity in~\eqref{phi0lin} near the past boundary. Following Remark~\ref{rembigbang},
in a cosmological situation  this involves the need
to model the behavior of the solutions near in the early universe (for example near the big bang singularity).

We finally note that the solutions of the dynamical wave equation
satisfy causality and finite speed of propagation, similar as worked out in~\cite{dirac}.
The methods of the present paper do not give any new insights into causal propagation.
The same is true for the Cauchy problem.

\section{Perturbative Treatment of the Coupling Terms} \label{seccouple}
In the previous sections we constructed solutions of the equation~$\mathcal{Q} \psi = 0$
in a given time strip. In order to treat the errors caused by the coupling terms in Proposition~\ref{prpsecond},
we need to solve the inhomogeneous equation
\beq \label{Qinhom}
\mathcal{Q} \psi = \phi \:,
\eeq
where~$\phi$ is formed of the second variations of both~\eqref{D2lfe} and~\eqref{d2R}.
This equation is accompanied by a corresponding inhomogeneous bosonic equation.
The inhomogeneous equation~\eqref{Qinhom} can be solved in the time strip
by multiplying with the inverse operator~$\mathcal{Q}^{-1}$
(here we need to assume that the operator~$\mathcal{Q}$ is invertible or that
all inhomogeneities are admissible in the sense of Definition~\ref{defadmissible}).
Likewise, the inhomogeneous bosonic equation can be solved with the methods developed in~\cite{nonlocal}.
Iterating this procedure, one gets a desired solution of the linearized field equations,
provided that the iteration converges. Convergence can be shown if the inhomogeneity
obtained in each iteration step is sufficiently small.

Here we do not enter the details of this construction, but merely specify what we mean by ``small''.
In the dynamical wave equation we consider variations of individual physical wave functions
as computed in Proposition~\ref{prpsecond}. The contributions~\eqref{D2lfe} and~\eqref{d2R}
are in general singular on the light cone and diverge as~$\varepsilon \searrow 0$,
whereas~\eqref{D2Q} remains finite in this limit. In order to satisfy the linearized field equations,
all these singular contributions must cancel each other. Doing so leads to solving the bosonic
equations, as worked out in detail in~\cite{nonlocal}. Once these equations have been satisfied,
we can take the limit~$\varepsilon \searrow 0$. Then we can clearly no longer argue with the
smallness of~$\varepsilon$ (except for the parameter~$\kappa$ in~\eqref{d2L}). 
But there is another argument which shows that both~\eqref{D2lfe}
and~\eqref{d2R} are small compared to~\eqref{D2Q}:
The variation of~\eqref{D2lfe} and~\eqref{d2R} involves two factors of~$\delta P$
and thus four wave functions in total. Thus we obtain symbolically
\[ R(x,y) \simeq |\psi|^2\, |\delta \psi|^2 \:. \]
Since the wave functions are normalized (in the sense that their spatial integral is one),
the factor~$|\psi|^2$ scales like one over the spatial volume~$V$
in which the wave function is localized. Using that, in typical situations, the wave function
is localized on a length scale which is much larger than the Compton scale, we obtain
\[ |\psi|^2 \sim \frac{1}{V} \ll m^3 \:. \]
In this sense, the coupling of~\eqref{D2lfe} and~\eqref{d2R} to the dynamical wave equations
are very small and can thus be treated perturbatively.

\appendix
\section{Example: A Regularized Dirac Dynamics} \label{appex}
From the perspective of partial differential equations (PDEs), it might seem unusual or even
surprising that the dynamical wave equation is obtained by {\em{minimizing}} the causal action.
In contrast, hyperbolic equations like the scalar wave equation or the Dirac equation
are obtained typically by seeking for critical points of an action, whereas a minimization process
(like minimizing the Dirichlet energy)
typically gives rise to elliptic equations.
In order to explain and clarify this point, we now consider an example of a positive action
with the property that minimizing this action gives rise to EL equations being hyperbolic PDEs.
This example has a similar analytic structure as the effective action of the dynamical wave
equation~\eqref{SQ}, so much so that it can be regarded as the PDE analog of this action.

We consider the following action in Minkowski space,
\beq \label{Sactex}
S(\psi) := \int_{-\infty}^\infty dt \int_{-\infty}^\infty dt' 
\int_{\R^3} \!\!\!d^3x \; \Sl \big( (i \Pdd - m) \psi\big)(t, \vec{x}) \:|\: \gamma^0\: \eta(t-t')\:
\big( (i \Pdd - m) \psi\big)(t', \vec{x}) \Sr \:.
\eeq
Here~$\eta \in C^\infty(\R)$ is a symmetric cutoff function (i.e.~$\eta(t)=\eta(-t)$). We assume that its Fourier transform~$\hat{\eta}$ 
is strictly positive, i.e.\
\beq \label{strictpositive}
\hat{\eta}(\omega) > 0 \qquad \text{for all~$\omega \in \R$}\:. 
\eeq
As a concrete example, one may choose~$\eta$ as a Gaussian,
\[ \eta(\tau) = e^{-\tau^2 \delta} \qquad \text{with} \qquad \delta>0\:. \]
Considering first variations with compact support in spacetime, a direct computation gives
the corresponding EL equations
\beq \label{ELex}
(i \Pdd - m) \,\gamma^0\, (i \Pdd - m) \psi = 0 \:.
\eeq

\begin{Lemma} \label{lemmaE1}
The general solution~$\psi \in C^\infty_\sc(M, SM)$ of the EL equations~\eqref{ELex}
can be written as
\beq \label{psi12}
\psi(t,\vec{x}) = \psi_1^D(t,\vec{x}) + t\: \psi_2^D(t, \vec{x}) \:,
\eeq
where~$\psi_{1\!/\!2}^\text{D}$ are solutions of the Dirac equation,
\[ (i \Pdd - m) \,\psi_{1\!/\!2}^D = 0 \:. \]
\end{Lemma}
\Proof Taking the Fourier transform of~\eqref{ELex}, we obtain the algebraic equation
\begin{align}
0 &= \gamma^0\, (\slashed{k} - m) \,\gamma^0\, (\slashed{k} - m) \,\hat{\psi}(k) \label{dirmat} \\
&= \gamma^0\, \big\{ \slashed{k},  \gamma^0 \big\}\, (\slashed{k} - m) \,\hat{\psi}(k) 
+(\gamma^0)^2 \:(-\slashed{k} - m) \,(\slashed{k} - m) \,\hat{\psi}(k) \\
&= 2 k^0\, \gamma^0\,\, (\slashed{k} - m) \,\hat{\psi}(k) 
- (k^2-m^2) \,\hat{\psi}(k) \\
&= \big( |k^0|^2 + |\vec{k}|^2 + m^2 - 2\,k^0 \gamma^0\,\vec{k} \vec{\gamma} 
- 2 m\, k^0 \gamma^0 \big)\, \hat{\psi}(k) \:. \label{matrix}
\end{align}
The calculation
\[ \big( - 2\,k^0 \gamma^0\,\vec{k} \vec{\gamma} - 2 m\, k^0 \gamma^0 \big)^2
= 4\, |k^0|^2\: \big( \gamma^0\,\vec{k} \vec{\gamma} +  m \gamma^0 \big)^2
= 4\, |k^0|^2\: \big( |\vec{k}|^2 + m^2 \big) \]
shows that the matrix in~\eqref{dirmat} has the eigenvalues
\[  |k^0|^2 + |\vec{k}|^2 + m^2 \pm 2 k_0\, \sqrt{ |\vec{k}|^2 + m^2 }
= \Big( |k^0| \pm \sqrt{ |\vec{k}|^2 + m^2 } \Big)^2 \:. \]
In particular, one sees that the eigenvalues vanish only on the mass shells. Moreover, 
the $k^0$-derivative of the matrix in~\eqref{dirmat} is singular also only on the mass shells, because
\begin{align*}
\frac{\partial}{\partial k^0} \gamma^0\, (\slashed{k} - m) \,\gamma^0\, (\slashed{k} - m)
= 2 \gamma^0\, (\slashed{k} - m) \:.
\end{align*}
Therefore, the only distributional solutions of~\eqref{matrix} have the form
\[ \hat{\psi}_1(k)\: \delta\big( k^2 - m^2 \big) \qquad \text{and} \qquad
\frac{\partial}{\partial k^0} \Big( \hat{\psi}_2(k)\: \delta\big( k^2 - m^2 \big) \Big) \:. \]
Transforming back to position space gives the result.
\QED

We next restrict the above action to wave functions~$\psi \in C^\infty_0( (-t_{\min}, t_{\max}) \times \R^3)$
supported in a time strip. This means in particular that we are not allowed to insert the solutions~\eqref{psi12}
into the action. But we may introduce a smooth cutoff function~$\theta$ and take the functions
\[ \psi(t,\vec{x}) = \theta(t)\: \big( \psi_1^D(t,\vec{x}) + t\: \psi_2^D(t, \vec{x}) \big) \qquad \text{with} \qquad
\theta \in C^\infty_0\big( (-t_{\min}, t_{\max}) \big) \:. \]
In the next lemma we verify that our action is strictly positive on such wave functions.
\begin{Lemma} \label{lemmaE2} Let~$\psi$ be a smooth solution of the EL equations~\eqref{ELex}
of the form~\eqref{psi12} defined in all of Minkowski space. Then the action~\eqref{Sactex} takes the form
\begin{align}
&S\big( \theta\: (\psi_1^D + t\, \psi_2^D ) \big) = \int_{-\infty}^\infty dt \int_{-\infty}^\infty dt' \int_{\R^3} \!\!\!d^3x
\notag \\
&\times \Sl  \Big( \theta'(t)  (\psi_1^D + t\, \psi_2^D ) + \theta(t)\, \psi_2^D \Big) \notag \\
&\qquad\qquad \qquad \times \:|\: \gamma^0\: \eta(t-t')\:
\Big( \theta'(t')  (\psi_1^D + t'\, \psi_2^D ) + \theta(t')\, \psi_2^D \Big) \Sr \:. \label{Stheta}
\end{align}
This is strictly positive unless~$\psi$ vanishes.
\end{Lemma}
\Proof Noting that
\[ (i \Pdd - m) \,\Big( \theta(t)  (\psi_1^D + t\, \psi_2^D ) \Big)
= i \gamma^0 \Big( \theta'(t)  (\psi_1^D + t\, \psi_2^D ) + \theta(t)\, \psi_2^D \Big) \:, \]
we obtain~\eqref{Stheta}.
The strict positivity statement follows immediately by taking the Fourier transform in time
and using~\eqref{strictpositive}.
\QED

Integrating by parts, the above action~\eqref{Sactex} can be written in the form~\eqref{SQ} if we introduce the distributional
kernel
\beq \label{Qexform}
Q(x,y) =  (i \Pdd_x - m)\, \gamma^0\,  (-i \Pdd_y - m) \,\Big( \eta(t-t')\: \delta^3(\vec{x}-\vec{y}) \Big)\:.
\eeq
Now we can use~\eqref{cip} to introduce a corresponding conserved inner product.
\begin{Lemma} \label{lemmaE3}
Let~$\psi$ and~$\phi$ be solutions of the EL equations~\eqref{ELex}. Then
\begin{align*}
\la \psi | \phi \ra^{t_2}
&= -\int_M d^4x \int_M d^4y\:\eta(t-t')\:\Big( \delta(t-t_2) + \delta(t'-t_2) \Big)\: \delta^3(\vec{y}-\vec{x}) \\
&\qquad\: \times \Big(\Sl \psi(x) \:|\: (i \Pdd_y-m) \phi(y) \Sr + \Sl (i \Pdd_x-m) \psi(x) \:|\: \phi(y) \Sr  \Big) \:.
\end{align*}
\end{Lemma}
\Proof For notational simplicity, we compute the commutator inner product at time zero, so that
\begin{align*}
\la \psi | \phi \ra^\Omega = -2i \int_M d^4x \int_M d^4y\: \Big( \Theta(-t)\, \Theta(t')
- \Theta(t)\, \Theta(-t') \Big) \:\Sl \psi(x) \:|\: Q(x,y)\, \phi(y) \Sr \:.
\end{align*}
Inserting~\eqref{Qexform} and integrating by parts, we obtain
\begin{align*}
\la \psi | \phi \ra^\Omega &= i \int_M d^4x \int_M d^4y\:\eta(t-t')\:\delta^3(\vec{y}-\vec{x}) \\
&\quad\: \times \Big( \Sl (i \Pdd_x-m) \big( \Theta(-t)\, \psi(x)\big) \:|\: \gamma^0\, (i \Pdd_y-m) \big(\Theta(t')\, \phi(y) \big) \Sr \\
&\qquad\qquad\;\: - \Sl (i \Pdd_x-m) \big( \Theta(t)\, \psi(x)\big) \:|\: \gamma^0\, (i \Pdd_y-m) \big(\Theta(-t')\, \phi(y) \big) \Sr \Big) \\
&= i \int_M d^4x \int_M d^4y\:\eta(t-t')\:\delta^3(\vec{y}-\vec{x}) \\
&\quad\: \times \Big( i \delta(t)\,\Sl \psi(x)\big) \:|\: (i \Pdd_y-m) \phi(y) \Sr \\
&\qquad\quad+\Theta(-t)\: \Sl (i \Pdd_x-m) \psi(x) \:|\: \gamma^0\, (i \Pdd_y-m) \big(\Theta(t')\, \phi(y) \big) \Sr \\
&\qquad\quad - \Theta(t)\: \Sl (i \Pdd_x-m)\psi(x) \:|\: \gamma^0\, (i \Pdd_y-m) \big(\Theta(-t')\, \phi(y) \big) \Sr \Big)
\:.
\end{align*}
Now we integrate the Dirac operator by parts from the right to the left and use the relation
\[ \Big( \frac{d}{dt} + \frac{d}{dt'} \Big) \eta(t-t') = 0 \]
as well as the EL equations~\eqref{ELex}. This gives
\begin{align*}
&\la \psi | \phi \ra^\Omega \\
&= i \int_M d^4x \int_M d^4y\:\eta(t-t')\:\delta^3(\vec{y}-\vec{x}) \;
\Big( i \delta(t)\,\Sl \psi(x)\big) \:|\: (i \Pdd_y-m) \phi(y) \Sr \\
&\qquad\quad+i \delta(t)\: \Sl (i \Pdd_x-m) \psi(x) \:|\: \Theta(t')\, \phi(y) \Sr +i \delta(t)\: \Sl (i \Pdd_x-m)\psi(x) \:|\: \Theta(-t')\, \phi(y) \Sr \Big) \\
&= i \int_M d^4x \int_M d^4y\:\eta(t-t')\:\delta^3(\vec{y}-\vec{x}) \\
&\quad\: \times \Big( i \delta(t)\,\Sl \psi(x)\big) \:|\: (i \Pdd_y-m) \phi(y) \Sr +i \delta(t)\: \Sl (i \Pdd_x-m) \psi(x) \:|\: \phi(y) \Sr  \Big) \\
&= -\int_M d^4x \int_M d^4y\:\eta(t-t')\:\delta(t)\: \delta^3(\vec{y}-\vec{x}) \\
&\qquad\: \times \Big(\Sl \psi(x)\big) \:|\: (i \Pdd_y-m) \phi(y) \Sr + \Sl (i \Pdd_x-m) \psi(x) \:|\: \phi(y) \Sr  \Big) \:.
\end{align*}
Clearly, the commutator inner product is invariant under complex conjugation and exchanging~$\psi$ with~$\phi$.
Therefore, we may symmetrize the right side as well. This gives the result.
\QED

Using the representation~\eqref{psi12} of the solution, we obtain
\begin{align*}
\la \psi | \phi \ra^{t_2}
&= -\int_M d^4x \int_M d^4y\:\eta(t-t')\:\Big( \delta(t-t_2) + \delta(t'-t_2) \Big)\: \delta^3(\vec{y}-\vec{x}) \\
&\qquad\: \times \Big(\Sl \psi(x) \:|\: i \gamma^0 \phi_2^D(y) \Sr + \Sl i \gamma^0 \psi_2^D(x) \:|\: \phi(y) \Sr 
\Big) \\
&= -i \int_M d^4x \int_M d^4y\:\eta(t-t')\:\Big( \delta(t-t_2) + \delta(t'-t_2) \Big)\: \delta^3(\vec{y}-\vec{x}) \\
&\qquad\: \times \Big(\Sl \psi(x) \:|\: \gamma^0 \phi_2^D(y) \Sr + \Sl \psi_2^D(x) \:|\: \gamma^0 \phi(y) \Sr \Big) \\
&= -i \int_M d^4x \int_M d^4y\:\eta(t-t')\:\Big( \delta(t-t_2) + \delta(t'-t_2) \Big)\: \delta^3(\vec{y}-\vec{x}) \\
&\qquad\: \times \Big(\Sl \psi_1^D(x) \:|\: \gamma^0 \phi_2^D(y) \Sr + \Sl \psi_2^D(x) \:|\: \gamma^0 \phi_1^D(y) \Sr \Big) \\
&= -i \big( ( \psi_1^D \:|\: \eta * \phi_2^D ) + ( \psi_2^D \:|\: \eta * \phi_1^D )
- ( \eta * \psi_1^D \:|\: \phi_2^D ) + ( \eta*\psi_2^D \:|\:  \phi_1^D ) \big) \:,
\end{align*}
where the star denotes convolution in time and~$(.|.)$ is the usual $L^2$-scalar product on Dirac spinors.
Noting that the convolution preserves the Dirac equation, this is conserved in time, as desired.

We finally discuss the above example and explain the similarities and differences to the
dynamical wave equation. We point out
that the matrix in~\eqref{matrix} involves a bilinear contribution and thus does not have a vector-scalar
structure, in contrast to the action of the dynamical wave equation~\eqref{SQ} when computed
in Minkowski space (for details see~\cite{dirac, noether}). This can be regarded as being a
shortcoming of the ansatz~\eqref{Sactex}. This shortcoming can be avoided by choosing more
sophisticated ans\"atze. An additional feature which one can build in is that the action is
Lorentz invariant except for the regularization effects. For brevity, here we shall not enter a
discussion of the different alternative ans\"atze, but restrict attention to the simple ansatz~\eqref{Sactex}.

The main motivation for the action~\eqref{Sactex} comes from the fact
it is positive (see Lemma~\ref{lemmaE2}), making it possible to minimize the action.
Nevertheless, the resulting EL equations give rise to {\em{hyperbolic}} equations (see Lemma~\ref{lemmaE1}).
Just as explained for the dynamical wave equation in Section~\ref{secsoldyn}, the prize to pay is that
the resulting conserved inner product is not positive definite (see Lemma~\ref{lemmaE3}).
Therefore, following the procedure in Section~\ref{secpositive}, in order to introduce a
Hilbert space of solutions, one needs to choose a positive definite subspace of the whole solution space.
This can be accomplished by choosing the subspace spanned by solutions of the form
\[ \psi(t,\vec{x}) = \psi^D(t,\vec{x}) + i \lambda t\: \psi^D(t, \vec{x}) \]
with~$\psi^D$ a solution of the Dirac equation. Here~$\lambda>0$ is a small parameter
(similar to the perturbation parameter~$\lambda$ in Section~\ref{secpositive}).
Then, linearly in~$\lambda$, the inner product in Lemma~\ref{lemmaE3} simplifies to
\begin{align*}
\la \psi | \psi \ra^{t_2}
&= -\int_M d^4x \int_M d^4y\:\eta(t-t')\:\Big( \delta(t-t_2) + \delta(t'-t_2) \Big)\: \delta^3(\vec{y}-\vec{x}) \\
&\qquad\qquad\qquad\qquad\qquad\qquad\: \times \Sl \psi^D(x) \:|\: \gamma^0 \psi^D(y) \Sr + \O \big( \lambda^2 \big)\:.
\end{align*}
This is positive, as desired. Indeed,taking for~$\eta$ a Dirac sequence and passing to the limit,
one recovers the usual scalar product on Dirac solutions.

\section{Supplementary Material and Related Considerations} \label{appsupplement}
The constructions in the present paper evolved in various steps from the earlier constructions in~\cite{dirac}.
In this appendix, we review a few considerations which pointed towards shortcomings of the previous
constructions and gave ideas for how to overcome them.
Presenting these considerations here serves the purpose of clarifying the main constructions of the
present paper, putting them into a somewhat broader context and thereby conveying a complete and
coherent picture of the structure of the causal action principle.
We begin by explaining the shortcomings of the construction of the extended Hilbert space
in~\cite{dirac} (Appendix~\ref{appdirac}). Then we explain why, 
consistent with earlier assumptions (for example in~\cite{dirac, fockfermionic, intro}),
the commutator inner product can coincide with the Hilbert space scalar product~$\la .|.\ra_\H$ only on
a proper subspace of~$\H$ (Appendix~\ref{appA}).
Moreover, we explain in Appendix~\ref{appMP}
why the assumption of a {\em{distributional ${\mathcal{M}}P$-product}}
as first introduced in~\cite[\S5.6]{pfp} and used in~\cite{reg, vacstab} cannot hold.
From today's perspective, this assumption, which looks appealing and reasonable at first sight,
is too simple and naive.
Next, in Appendix~\ref{appregQ} it is shown that, in Minkowski space, the Fourier transform~$\hat{Q}(k)$
of the integral kernel~$Q(x,y)$ in appearing in the first variation of the Lagrangian~\eqref{delLintro}
is indeed differentiable. This result shows that the computation
in~\cite[Section~5]{noether} which relates the commutator product to a discontinuity of the
semi-derivatives of~$\hat{Q}$ on the mass shell is incomplete.
This analysis is superseded by the present result in Section~\ref{secpositive} which relates the
positivity of the commutator inner product not to the form of~$\hat{Q}(k)$ at present, but instead to
boundary conditions in the past. Finally, in Appendix~\ref{appD} we analyze what finite propagation
speed tells us about the form of~$\hat{Q}(k)$.

\subsection{Shortcomings of the Previous Construction of the Extended Hilbert Space} \label{appdirac}
As already mentioned in the introduction, the 
construction of the extended Hilbert space in~\cite{dirac} has the shortcoming that
it is not manifestly canonical. More precisely, this can be from the fact that 
the construction it is based on linear perturbations of the system,
parametrized by so-called {\em{compatible generators}} (see~\cite[Definition~5.2]{dirac}).
The resulting compatibility conditions are shown to be solvable (see~\cite[Section~5.2]{dirac}).
But it is not clear in general if the solutions are unique. This means that, depending on which generators are
chosen, one might get different extended Hilbert spaces with potentially different dynamics.
Clearly, this situation is not satisfying from the conceptual point of view.
Therefore, it is an important task to improve the construction of the extended Hilbert space
in such a way that it becomes independent of the choices of
compatible generators. It is the main objective of the present paper to provide such a canonical construction.

\subsection{Why the Commutator Inner Product Cannot Represent $\la .|. \ra_\H$ on the Whole Hilbert Space}
\label{appA}
The critical reader may wonder why in Definition~\ref{defSLrep}
we restricted attention to a subspace~$\H^\fermi \subset \H$.
The reason is that it does not seem sensible to assume that the relation~\eqref{Ccond}
holds on the whole Hilbert space. For technical simplicity, we give the argument only in the
finite-dimensional setting.
\begin{Prp} Assume that~$\H$ is finite-dimensional and that~$c \neq 0$.
Then the equation~\eqref{Ccond} cannot hold for all~$u, v \in \H$.
\end{Prp}
\Proof
Let us assume conversely that~\eqref{Ccond} does hold on the whole Hilbert space, i.e.\
\beq \label{Ccondtot}
\la u|u \ra^\Omega_\rho = c\, \la u|u \ra_\H \qquad \text{for all~$u \in \H$ and~$c \neq 0$} \:.
\eeq
In order to get a contradiction, we first rewrite the commutator inner product as
\begin{align}
\la u|u \ra^\Omega_\rho &= \gamma^{\Omega}_\rho(\Comm) 
= \int_\Omega d\rho(x) \int_{M \setminus \Omega} d\rho(y) \: \big(D_{1,\Comm} - D_{2, \Comm}) \L(x,y) \notag \\
&= \bigg( \int_\Omega d\rho(x) \int_{M \setminus \Omega} d\rho(y)
- \int_{M \setminus \Omega} d\rho(x) \int_\Omega d\rho(y) \bigg) \: D_{1,\Comm} \L(x,y) \:, \label{commint}
\end{align}
where~$\Comm$ is the commutator-jet
\[ \Comm(x) := i \big[\scrA, x \big] \qquad \text{with} \qquad 
\scrA := |u \ra \la u | \]
(i.e.\ $\scrA \psi := \la u | \psi\ra_\H \: u$). Being linear in the commutator jet, the variational derivative
can be written as
\begin{align*}
D_{1,\Comm} \L(x,y) &= \tr_\H \big( \Comm(x)\, D_{1} \L(x,y) \big) =  i \:\tr_\H \big(  \big[\scrA, x \big]\, D_{1} \L(x,y) \big) \\
&=  i \:\tr_\H \big(  \scrA\, \big[x,  D_{1} \L(x,y) \big]\big)
=  i \:\Big\la u \,\Big|\,\big[x,  D_{1} \L(x,y) \big]\, u \Big\ra_\H \:.
\end{align*}
Using this relation in~\eqref{commint}, we obtain
\beq \label{comm1}
\la u|u \ra^\Omega_\rho =  i \int_M \big\la u \,\big|\, [x,  B(x)]\, u \big\ra_\H\: d\rho(x) \:,
\eeq
where the operator~$B(x)$ is defined by
\[ B(x) = \chi_\Omega(x) \int_{M \setminus \Omega} D_{1} \L(x,y) \: d\rho(y) 
- \chi_{M \setminus \Omega}(x) \int_\Omega D_{1} \L(x,y) \: d\rho(y) \:. \]
Finally, we rewrite~\eqref{comm1} as
\[ \la u|u \ra^\Omega_\rho = \la u \,|\,C u \ra_\H \]
with
\beq \label{Cdef}
C := i \int_M [x,  B(x)] \: d\rho(x) \:.
\eeq
If~\eqref{Ccondtot} holds, the operator~$C$ is a multiple of the identity,
\[ C = c\, \1_\H \:. \]
As a consequence, the operator~$C$ has a non-zero trace.
On the other hand, taking the trace of~\eqref{Cdef} and using the 
commutator structure, we get zero. This is a contradiction.
\QED

\subsection{Why the Assumption of a Distributional $\mathcal{M} P$-Product Cannot Hold} \label{appMP}
The assumption of a distributional $\mathcal{M} P$-product was introduced in~\cite[Defintion~5.6.3]{pfp}
and used in the analysis of regularizations in~\cite{reg}.
It is based on the observation that the kernel~$Q(x,y)$ describing the first variations of the
Lagrangian~\eqref{delLdef} can be factorized into the product
\beq \label{MPpos}
Q(x,y) = \frac{1}{2}\: {\mathcal{M}}(x,y)\: P(x,y) \:,
\eeq
where~${\mathcal{M}}(x,y)$ is the first variation of the Lagrangian, regarded as a function
of the closed chain~$A=A_{xy}$ in~\eqref{Axy}; more precisely,
\beq \label{Mdef}
{\mathcal{M}}(x,y)^\alpha_\beta :=
\frac{\partial {\mathcal{L}}[A]}{\partial A^\beta_\alpha} \bigg|_{A = A_{xy}}
\eeq
(for the derivation of~\eqref{MPpos} see~\cite[Lemma~5.2.1]{pfp}).
In the Minkowski vacuum, all the kernels only depend on the difference vector~$\xi:=y-x$.
Taking the Fourier transform, the product~\eqref{MPpos} becomes a convolution, i.e.\
\beq \label{MPmom}
\hat{Q} (q) = \frac{1}{2} \:\int \frac{d^4p}{(2\pi)^4}\:
\hat{\mathcal{M}} (p) \:\hat{P}(q-p)\:.
\eeq

\begin{figure}[tb]
\begin{center}
\scalebox{0.9}
{\includegraphics{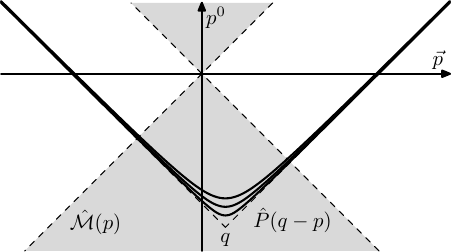}}
\caption{The convolution~${\hat{\mathcal{M}}} * \hat{P}$.}
\label{fig4}
\end{center}
\end{figure}

These formulas hold for the {\em{regularized}} objects. This means that we must first
regularize~$P(x,y)$ and then compute~${\mathcal{M}}(x,y)$ via~\eqref{Mdef}.
The {\em{assumption of a distributional ${\mathcal{M}}P$-product}}
states that in~\eqref{MPpos} one may take the product with the {\em{un}}regularized kernel~$P(x,y)$.
This assumption is motivated by the structure of the corresponding convolution integral~\eqref{MPmom},
as we now explain (for a detailed account of this point see~\cite[Section~2.2]{reg}).
Away from the light cone (i.e.\ if~$\xi^2 \neq 0$), one can compute~${\mathcal{M}}(x,y)$
without regularization. It is a smooth function which vanishes if~$\xi$ is spacelike and is anti-symmetric,
i.e.\ ${\mathcal{M}}(x,y) = -{\mathcal{M}}(y,x)$.
Clearly, on the light cone the regularization must be taken into account, making the computations more
difficult and less explicit. In order to keep the setting as simple as possible,
in~\cite[Section~5.6]{pfp} it is assumed that~${\mathcal{M}}(x,y)$ is well-defined as a distribution if the
regularization is removed. Moreover, it is assumed that the resulting distribution is again anti-symmetric.
Under these assumptions, its Fourier transform~$\hat{\mathcal{M}}(p)$ is supported inside the
mass cone, as depicted in Figure~\ref{fig4} as the gray shaded region.
As a consequence, the convolution~\eqref{MPmom} is well-defined and finite if~$q$ is chosen
in the lower mass cone, because the integrand vanishes outside a bounded set
(see Figure~\ref{fig4}, where three Dirac seas are shown).
However, if~$q$ is chosen outside the lower mass cone, then the integration range in~\eqref{MPmom}
is unbounded, and the integral will typically diverge.
The assumption of a distributional ${\mathcal{M}}P$-product states that for~$q$ in the lower
mass cone, the distribution~$\hat{Q}(q)$ (defined by~\eqref{MPmom} and removing the regularization)
can be computed with this procedure.

Before going on, we point out that the assumption of a distributional ${\mathcal{M}}P$-product
is not of fundamental significance. Instead, it is merely a simplifying assumption for the analysis of
the causal action principle in the vacuum. At the time, it seemed appealing also because
the geometric picture in Figure~\ref{fig4} seemed to give a simple explanation for the fact that
in the vacuum the one-particle states of negative energy (and not the states of positive energy
or the whole solution space) are occupied.
A drawback of this assumption is that~$\hat{Q}$ diverges in the upper mass cone.
This means that, if one wants to recover the positive energy solutions of the Dirac equation
as solutions of the linearized field equations~\eqref{lfehom}, then the term~$Q\, \delta \Psi$
diverges if the regularization is removed. In~\cite{dirac}, this problem was bypassed
by assuming that this divergence is compensated by the first summand~$(DQ|_{\Psi}(\delta \Psi)) \,\Psi$
in~\eqref{lfehom} (this leads to introducing the kernel~$Q^\text{dyn}$; for details see
\cite[Sections~5.3 and~5.4]{dirac}).
However, as we will show below, a compensation of contributions between the first and second
summand in~\eqref{lfehom} contradicts the
approximate decoupling of the linearized field equations as worked out in Section~\ref{secdecouple}.
Therefore, the following considerations show that 
the assumption of a distributional ${\mathcal{M}}P$-product is too naive in the sense that
it cannot hold for causal fermion systems describing our universe.
As a consequence, some of the constructions in~\cite{dirac} are superseded by the
constructions in the present paper. Moreover, we learn that those results in the papers~\cite{reg, vacstab},
where the assumption of a distributional ${\mathcal{M}}P$-product was used, need to be modified
before they apply to physically realistic situations.

We now give the argument which rules out
the assumption of a distributional ${\mathcal{M}}P$-product.
Assume conversely that the assumption of a distributional ${\mathcal{M}}P$-product holds.
Then, using that the integration range of the convolution in momentum space is compact
(see Figure~\ref{fig4}), the operator~$\hat{Q}(k)$ is well-defined and finite inside the lower mass cone.
Consequently, evaluating~\eqref{ELQ} for a state on the lower mass shell, one concludes that
the parameter~$\mathfrak{r}$ is also finite.
However, $\hat{Q}(k)$ diverges inside the upper mass cone, with a singularity of the form
(see~\cite{vacstab})
\beq \label{hQvac}
\hat{Q}(k) \simeq \int \frac{m^3}{\xi^4}\: \slashed{\xi}\: P(x,y) \: d^4 \xi \simeq \frac{m^3}{\varepsilon^2}
\eeq
(here $\simeq$ means equality up to an irrelevant numerical constant).
In order to make contact to relativistic quantum mechanics, the linearized field equations must
admit solutions in the upper mass cone. Consequently, the left side of~\eqref{diverge}
has a singularity~$\sim \varepsilon^{-2}$. The term on the right side of~\eqref{diverge}, however,
diverges only~$\sim \varepsilon^{-1}$, as we now explain.
According to~\eqref{Rdef}, we must consider the two contributions from~\eqref{R1} and~\eqref{R2}.
In~\eqref{R2} the Lagrange parameter~$\kappa$ of the boundedness constraint appears.
As shown in~\cite[Appendix~A]{jacobson} it has the scaling
\[ 
\kappa \lesssim (\varepsilon m)^p + \Big( \frac{\varepsilon}{\delta} \Big)^{8-\hat{s}} \:. \]
with~$p \geq 4$ and~$\hat{s} \in \{0,2\}$. Taking the Fourier transform gives a contribution with the scalings
\[ \hat{Q}(k) \lesssim (\varepsilon m)^4\: \frac{1}{\varepsilon^5} = \frac{m^4}{\varepsilon} \:, \]
which differs from~\eqref{hQvac} by an additional factor~$\varepsilon$, as desired.

It remains to consider the contribution~\eqref{R1}. Here the first factor involves a scaling factor~$\sim m^3$,
because the eigenvalues have the same absolute values up to order~$\O(m^3)$.
Next, we need to consider the two variational derivatives. One of them is used for testing; it appears
similarly when varying~\eqref{D2Q}. The second variational derivative, however, describes the
linear perturbation of the system. The strength of this perturbation gives another scaling
factor~$\varepsilon/\ell_\text{macro}$ or~$\varepsilon m$. Taking again the Fourier transform, we obtain the scaling
\[ \hat{Q}(k) \lesssim (\varepsilon m)^3\: \frac{1}{\varepsilon^5} \: \Big( \varepsilon m + \frac{\varepsilon}{\ell_\text{macro}} \Big) \:. \]
This again differs from~\eqref{hQvac} by a scaling factor~$\varepsilon$.

We conclude that the right side of~\eqref{diverge} is by a scaling factor~$\varepsilon$ smaller than the
left side. This is a contradiction, showing that the assumption of a distributional $\mathcal{M} P$-product
cannot hold.

\subsection{On the Regularity of~$\hat{Q}$} \label{appregQ}
In~\cite[Section~5.2]{noether} the commutator inner product was analyzed in Minkowski space. It was shown
that it agrees with the usual scalar product on Dirac wave functions, up to a prefactor which is
proportional to the discontinuity of the first derivative of~$\hat{Q}(k)$, i.e.\ to the expression
\beq \label{diffjump}
\partial_\omega^+ \hat Q( -\omega, \vec k) - \partial_\omega^- \hat Q( -\omega, \vec k) \:,
\eeq
where~$\omega := \sqrt{|\vec{k}|^2-m^2}$, and the semi-derivatives are defined by
\begin{align*}
\partial_\omega^+ \hat Q( -\omega, \vec k) &= \lim_{h \searrow 0}\:
\frac{1}{h}\, \Big( \hat Q( -\omega+h, \vec k) - \hat Q( -\omega, \vec k) \Big)  \\
\partial_\omega^- \hat Q( -\omega, \vec k) &= \lim_{h \nearrow 0}\:
\frac{1}{h}\, \Big( \hat Q( -\omega+h, \vec k) - \hat Q( -\omega, \vec k) \Big) \:.
\end{align*}
In view of this result, it is an important task to determine the regularity of~$\hat{Q}$ and to verify
whether this function can really have a discontinuity in its first derivative.
We now analyze this question. We will find that~$\hat{Q} \in C^1(\hat{M})$ is indeed continuously
differentiable, showing that the expression~\eqref{diffjump} is zero.
This also shows that the computation of the commutator inner product in~\cite[Section~5.2]{noether}
is incomplete. This shortcoming is overcome by the construction in Section~\ref{secpositive}.

In general terms, the regularity of~$\hat{Q}(k)$ in momentum space
is related to decay properties of~$Q(\xi)$ in position space. In order to see how this comes about,
let us begin with the model example of a Lorentz invariant, complex-valued distribution of the form
\beq \label{Rxy}
U(x,y) := \int \frac{d^4p}{(2\pi)^4}\: f(p^2)\: \Theta\big(-p^2 \big)\: e^{-ip(x-y)} \:.
\eeq
We consider the situation that for a parameter~$a_0>0$, the function~$f \in C^0_0(\R^+) \cap
C^\infty(\R^+ \setminus \{a_0\})$
is piecewise smooth and vanishes at~$p^2=a_0$ with a discontinuity of the first derivative, i.e.\
\[ \lim_{a \rightarrow a_0, \;a \neq a_0} f(a) = 0 \qquad \text{and} \qquad
c:= \lim_{a \searrow a_0} f'(a) - \lim_{a \nearrow a_0} f'(a) \neq 0 \:. \]
Clearly, the distribution~$U(x,y)$ depends only on the difference vector~$\xi := y-x$.
The behavior of the function~$U(x,y)$ for large~$\xi$ could be analyzed in detail using methods
of Fourier analysis. The result needed here can be seen most easily as follows.
We first rewrite the kernel~$U(x,y)$ as
 \beq \label{Rconv}
 U(x,y) := \int_0^\infty f(a)\: T_a(x,y)\: da \:,
\eeq
 where~$T_a(x,y)$ is the Fourier transform of the lower mass shell,
 \[ T_a(x,y) := \int \frac{d^4p}{(2\pi)^4}\: \delta(p^2-a)\: \Theta\big(-p^2 \big)\: e^{-ip(x-y)}\:. \]
This Fourier integral is a well-defined tempered distribution. It has a singular contribution on the
light cone (i.e.\ if~$(y-x)^2=0$), which is polynomial in~$a$, so that the integral in~\eqref{Rconv} is
well-defined, giving again a singular distribution on the light cone.
Away from the light cone, however, $T_a(x,y)$ is a smooth function given by
(see~\cite[eq.~(1.2.29)]{cfs})
\[ 
T_a(x,y) = \left\{ \begin{array}{cl} 
\displaystyle \frac{\sqrt{a}}{16 \pi^2} \:\frac{Y_1\big(\sqrt{a \xi^2} \,\big)}{\sqrt{\xi^2}}
+\frac{i \sqrt{m}}{16 \pi^2}\: \frac{J_1 \big(\sqrt{a \xi^2} \,\big)}{\sqrt{\xi^2}}\: \epsilon(\xi^0)
& \text{if~$\xi$ is timelike} \\[1em]
\displaystyle \frac{\sqrt{a}}{8 \pi^3} \frac{K_1 \big(\sqrt{-a \xi^2} \,\big)}{\sqrt{-\xi^2}} & \text{if~$\xi$ is spacelike}\:,
\end{array} \right. \hspace*{-0.3em} \]
where~$J_1$, $Y_1$ and~$K_1$ are Bessel functions. For large values of their arguments, these
Bessel functions have the asymptotics (see~\cite[eqs~10.7.8 and~10.25.3]{DLMF})
\begin{align*}
J_1(z) &= \sqrt{\frac{2}{\pi z}}\: \cos\Big( z - \frac{\pi}{2} -\frac{\pi}{4} \Big) + \O \big(z^{-\frac{3}{2}} \big) \\
Y_1(z) &= \sqrt{\frac{2}{\pi z}}\: \sin\Big( z - \frac{\pi}{2} -\frac{\pi}{4} \Big) + \O \big(z^{-\frac{3}{2}} \big) \\
K_1(z) &= \sqrt{\frac{\pi}{2z}}\: e^{-z} \;\Big( 1 + \O \big(z^{-1} \big) \Big) \:.
\end{align*}
Therefore, the function~$T_a$ decays exponentially in spacelike directions, whereas
for large timelike directions, it has the asymptotics
\beq \label{Taasy}
T_a(x,y) \simeq \frac{a^\frac{1}{4}}{\big( \xi^2 \big)^\frac{3}{4}}
\exp\Big(\pm i \sqrt{a \xi^2}\Big) \Big( 1 + \O \Big( \frac{1}{\sqrt{a \xi^2}} \Big) \Big)
\eeq
(here $\simeq$ again means equality up to an irrelevant numerical constant).
This formula holds similarly for derivatives of~$T_a$, and the resulting formulas are obtained
simply by differentiating the asymptotic formula. This makes it possible to determine the
asymptotics of~$U(x,y)$ with the following computation,
\begin{align}
U(x,y) &\simeq \frac{1}{\big( \xi^2 \big)^\frac{3}{4}} \int_0^\infty a^\frac{1}{4}\:f(a)\: \exp\Big(\pm i \sqrt{a \xi^2}\Big)\: da\; \Big( 1 + \O \Big( \frac{1}{\sqrt{a \xi^2}} \Big) \Big) \notag \\
&\simeq \frac{1}{\big( \xi^2 \big)^\frac{3}{4}} \int_0^\infty a^\frac{1}{4}\:f(a)\: \bigg( \frac{a}{\xi^2}\:
\frac{d^2}{da^2} \exp\Big(\pm i \sqrt{a \xi^2}\Big) \bigg)\: da\;
\Big( 1 + \O \Big( \frac{1}{\sqrt{a \xi^2}} \Big) \Big) \notag \\
&\simeq
\frac{a_0^\frac{1}{4}}{\big( \xi^2 \big)^\frac{3}{4}}\:\frac{a_0}{\xi^2} \:c
\exp\Big(\pm i \sqrt{a_0 \xi^2}\Big) \bigg)\: \Big( 1 + \O \Big( \frac{1}{\sqrt{a_0 \xi^2}} \Big) \Big) \:,
\label{Rxyasy}
\end{align}
where in the last step we integrated by parts twice and used that the boundary terms
at~$a=a_0$ arising from
the discontinuity of the first derivative of~$f$ give the leading contribution for large~$a \xi^2$.
Comparing with the asymptotics for~$T_{a_0}$ in~\eqref{Taasy}, we obtain the simple formula
\beq \label{Rsimp}
U(x,y) \simeq \frac{a_0}{\xi^2}\:c\: T_{a_0}(x,y) \:\Big( 1 + \O \Big( \frac{1}{\sqrt{a_0 \xi^2}} \Big) \Big) \:.
\eeq

The next question is how Dirac matrices change the picture. To this end, we modify~\eqref{Rxy}
by inserting a factor of~$\slashed{p}$,
\beq \label{rxy}
u(x,y) := \int \frac{d^4p}{(2\pi)^4}\: \slashed{p}\:f(p^2)\: \Theta\big(-p^2 \big)\: e^{-ip(x-y)} \:.
\eeq
This factor can be rewritten with the Dirac operator which can be pulled out of the integral, i.e.\
\[ u(x,y) = i \Pdd_x U(x,y) \:. \]
Consequently, differentiating~\eqref{Rxyasy} and comparing with the derivative of~\eqref{Taasy},
we conclude that
\begin{align}
u(x,y) &\simeq \frac{a_0^\frac{1}{4}}{\big( \xi^2 \big)^\frac{3}{4}}\:\frac{a_0}{\xi^2} \:c\:
\big( \sqrt{a}\: \slashed{\xi} \big)
\exp\Big(\pm i \sqrt{a_0 \xi^2}\Big) \bigg)\: \Big( 1 + \O \Big( \frac{1}{\sqrt{a_0 \xi^2}} \Big) \Big) \notag \\
&\simeq \frac{a_0}{\xi^2}\: (i \Pdd_x) T_a(x,y) \:. \label{rsimp}
\end{align}

Let us now analyze whether the asymptotics for large timelike~$\xi$ found in either~\eqref{Rsimp}
and~\eqref{rsimp} can be found in the operator~$Q(x,y)$.
Thus let~$\xi$ be a timelike vector. Since~$P(x,y)$ is smooth at~$\xi$, we may disregard the
regularization. It then follows by direct computation that
\begin{align*}
P_m(x,y) &= (i \Pdd_x + m)\, T_{m^2}(x,y) \\
&= \frac{m^\frac{3}{2}}{\big( \xi^2 \big)^\frac{3}{4}} \Big(\mp \frac{\slashed{\xi}}{\sqrt{\xi^2}} +\1 \Big)
\: \exp\Big(\pm i m\sqrt{\xi^2}\Big) \Big( 1 + \O \Big( \frac{1}{\sqrt{\xi^2}} \Big) \Big) \\
\mathcal{M}_{xy} &= P_m(x,y)\, P_m(y,x) - \frac{1}{4}\: \Tr \big( P_m(x,y)\, P_m(y,x) \big) \\
&\simeq \frac{m^3}{\big( \xi^2 \big)^\frac{3}{2}}\: \frac{\slashed{\xi}}{\sqrt{\xi^2}}
+ \O \big( \xi^{-4} \big) \\
Q(x,y) &= \mathcal{M}_{xy}\, P_m(x,y) \\
&\simeq \frac{m^3}{\big( \xi^2 \big)^\frac{3}{2}}\: \frac{\slashed{\xi}}{\sqrt{\xi^2}}\: 
\frac{m^\frac{3}{2}}{\big( \xi^2 \big)^\frac{3}{4}} \Big(\mp \frac{\slashed{\xi}}{\sqrt{\xi^2}} +\1 \Big)
\: \exp\Big(\pm i m\sqrt{\xi^2}\Big) \Big( 1 + \O \Big( \frac{1}{\sqrt{\xi^2}} \Big) \Big) \\
&\simeq \frac{m^3}{\big( \xi^2 \big)^\frac{3}{2}}\: P_m(x,y) \: \Big( 1 + \O \Big( \frac{1}{\sqrt{\xi^2}} \Big) \Big)\:.
\end{align*}
The main conclusion is that, in contrast to the prefactor~$1/\xi^2$ in~\eqref{Rsimp} and~\eqref{rsimp},
now a factor~$(\xi^2)^{-\frac{3}{2}}$ appears. Therefore, the kernel~$Q(x,y)$ decays faster at infinity,
meaning that its Fourier transform~$\hat{Q}$ is more regular than the integrands in~\eqref{Rxy}
and~\eqref{rxy}. We conclude that the discontinuity of the first derivatives in~$\hat{Q}$ in~\eqref{diffjump}
necessarily vanishes.

Here for simplicity we considered only one Dirac sea of mass~$m$. Considering several Dirac seas,
one also gets cross terms involving products of different masses. However, a direct computation shows
that the scaling factor~$(\xi^2)^{-\frac{3}{2}}$ remains unchanged. Therefore, the function~$\hat{Q}$ is
again differentiable on the mass hyperbolas.

\subsection{What Causal Propagation Speed Means for the Structure of~$\hat{Q}$} \label{appD}
Let us assume that, after removing the regularization, $\hat{Q}(p)$ is a well-defined, Lorentz invariant
distribution. Assuming vector-scalar structure, we can write it in the form
\beq \label{Qhatnoreg}
\hat{Q}(p) = f(p)\, \slashed{p} + g(p)
\eeq
where~$f$ and~$g$ are Lorentz invariant, real-valued functions.
The positivity statement of Proposition~\ref{prpQpos} implies that
\[ \left\{ \begin{array}{rl} f \sqrt{p^2} \geq |g| & \text{inside the upper mass cone} \\[0.3em]
-f  \sqrt{p^2} \geq |g| & \text{inside the lower mass cone}\:. \end{array} \right. \]
Outside the mass cone (i.e.\ if~$p^2<0$), the ansatz~\eqref{Qhatnoreg} is compatible with
positivity only if both~$f$ and~$g$ vanish. We conclude that~$\hat{Q}$ can be written as
\beq \label{Qform}
\hat{Q}(p) = \epsilon(p^0)\: \Theta(p^2)\: \times \left\{ \begin{array}{cl}
f^\vee(p^2)\, \slashed{p} + g^\vee(p^2) & \text{if~$p \in \mathcal{C}^\vee$} \\[0.3em]
f^\wedge(p^2)\, \slashed{p} + g^\wedge(p^2) & \text{if~$p \in \mathcal{C}^\wedge$}\:, \end{array} \right.
\eeq
where
\[ f^\vee \geq \big| g^\vee \big| \qquad \text{and} \qquad f^\wedge \geq \big| g^\wedge \big| \]
(and~$\mathcal{C}^\vee$ and~$\mathcal{C}^\wedge$ denote the upper respectively lower mass cone).

We now introduce the corresponding {\em{symmetric Green's operator}} as a distribution~$\hat{S}(p)$
with the properties
\[ \hat{S}(p)^* = \hat{S}(p) \qquad \text{and} \qquad \hat{Q}(p)\, \hat{S}(p) = \1 \quad
\text{for all~$p \in {\mathcal{C}}$}\:. \]
This distribution can be introduced by treating poles in the familiar way as principal value integrals
and distributional derivatives thereof.
Further restrictions on the form of~$\hat{S}$ are obtained by demanding causal propagation speed.
This is not a new physical input but is a consequence of the general structure of the causal action principle.
Moreover, causal propagation speed has been proven under general assumptions
using energy methods in~\cite{linhyp, dirac}.
In the setting considered here, causal propagation speed means that the distributional Fourier transform
of~$\hat{S}$ denoted by
\[ S(x,y) := \int \frac{d^4p}{(2 \pi)^4} \: \hat{S}(p)\: e^{-i p(x-y)} \]
must vanish if the difference vector~$y-x$ is spacelike.
This is indeed the case for symmetry reasons if
\beq \label{fgsymm}
f^\vee = f^\wedge \qquad \text{and} \qquad g^\vee = g^\wedge \:.
\eeq
The converse of this statement also holds under weak regularity and decay assumptions assumptions.
\begin{Lemma} \label{lemmaD1}
Assume that the distribution~$f := f^\vee - f^\wedge \in \D'(\R^+_0)$ has the following properties:
\begin{itemize}[leftmargin=2em]
\item[{\rm{(i)}}] It is regular and locally bounded except at a finite number of singular
points.
\item[{\rm{(ii)}}] It grows at most polynomially at~$a=0$ and~$a=\infty$.
\end{itemize}
Then the following implication holds: If the condition~\eqref{fgsymm} is violated, then there is
a spacelike vector~$\xi=y-x$ such that~$S(x,y) \neq 0$.
\end{Lemma}
\Proof Clearly, the distribution~$\hat{S}(p)$ is again Lorentz invariant. We write it
in analogy to~\eqref{Qform} as
\[ \hat{S}(p) = \epsilon(p^0)\: \Theta(p^2)\: \times \left\{ \begin{array}{cl}
\alpha^\vee(p^2)\, \slashed{p} + \beta^\vee(p^2) & \text{if~$p \in \mathcal{C}^\vee$} \\[0.3em]
\alpha^\wedge(p^2)\, \slashed{p} + \beta^\wedge(p^2) & \text{if~$p \in \mathcal{C}^\wedge$}\:, \end{array} \right. \]
with suitable distributions~$\alpha^\bullet$ and~$\beta^\bullet$ (which could have poles, as described above).
Its Fourier transform can be written as
\begin{align*}
S(x,y) &= \frac{1}{2} \int_0^\infty \bigg( \big(\alpha^\vee - \alpha^\wedge\big)(a)\: i \Pdd_x P_a(x,y)
+ \big(\alpha^\vee + \alpha^\wedge\big)(a)\: i \Pdd_x K_a(x,y) \bigg)\:da \\
&\quad\: +\frac{1}{2} \int_0^\infty \bigg( \big(\beta^\vee - \beta^\wedge\big)(a)\: P_a(x,y)
+ \big(\beta^\vee + \beta^\wedge\big)(a)\: K_a(x,y) \bigg)\:da \:,
\end{align*}
where~$P_a$ and~$K_a$ are the symmetric and anti-symmetric fundamental solutions of the Klein-Gordon
equation
\begin{align*}
P_a(x,y) &:= \int \frac{d^4p}{(2 \pi)^4} \: \delta(p^2-a)\: e^{-i p(x-y)} \\
K_a(x,y) &:= \int \frac{d^4p}{(2 \pi)^4} \: \delta(p^2-a)\: \epsilon(p^0)\: e^{-i p(x-y)} \:.
\end{align*}

Now suppose that~$\xi$ is spacelike. Then the anti-symmetric fundamental solution~$K_a$ vanishes.
Expressing the symmetric fundamental solution in position space in terms of the Bessel function~$K_1$,
\[ P_a(x,y) = -\frac{a}{4 \pi^3} \: \frac{K_1(\sqrt{-a \xi^2})}{\sqrt{-a \xi^2}} \]
(for details see for example~\cite[\S1.2.5]{cfs}), we obtain
\begin{align*}
S(x,y) &= -\frac{i}{8 \pi^3}\: (i \Pdd_x) \int_0^\infty \big(\alpha^\vee - \alpha^\wedge\big)(a)\:a\:
\frac{K_1(\sqrt{-a \xi^2})}{\sqrt{-a \xi^2}}\: da \\
&\quad\: -\frac{i}{8 \pi^3}\: \int_0^\infty \big(\beta^\vee - \beta^\wedge\big)(a)\:a\:
\frac{K_1(\sqrt{-a \xi^2})}{\sqrt{-a \xi^2}}\: da \\
&= -\frac{i}{4 \pi^3}\: (i \Pdd_x) \int_0^\infty \big(\alpha^\vee - \alpha^\wedge\big) \big(m^2 \big)\:m^2\:
\frac{K_1(m \sqrt{-\xi^2})}{\sqrt{-\xi^2}}\: dm \\
&\quad\: -\frac{i}{4 \pi^3}\: \int_0^\infty \big(\beta^\vee - \beta^\wedge\big) \big( m^2 \big)\:m^2\:
\frac{K_1(m \sqrt{- \xi^2})}{\sqrt{-\xi^2}}\: dm \:,
\end{align*}
where~$K_1$ is a modified Bessel function. In order to conclude that~$\alpha^\vee=\alpha^\wedge$
and~$\beta^\vee=\beta^\wedge$, one could argue with the invertibility of the Bessel or Mellin
transformations (as considered in the related paper~\cite{rooney}).
In order to weaken the regularity and decay assumptions,
we prefer to apply Lemma~\ref{lemmaD2} below. We conclude
that~$S$ vanishes in spacelike directions if and only if~$\alpha^\vee=\alpha^\wedge$
and~$\beta^\vee=\beta^\wedge$. Taking the Fourier transformation, we obtain the
corresponding symmetry property in position space~\eqref{fgsymm}.
\QED

\begin{Lemma} \label{lemmaD2}
Assume that~$f \in \D'(\R^+_0)$ is a distribution having the properties~{\rm{(i)}} and~{\rm{(ii)}}
in Lemma~\ref{lemmaD1}. Moreover, assume that the distributional integral
\[ g(\xi) := \int_0^\infty f(a)\: \frac{K_1(\sqrt{- a \xi^2})}{\sqrt{-\xi^2}}\: da \:. \]
vanishes for all spacelike~$\xi$. Then~$f$ is zero.
\end{Lemma}
\Proof Given any~$p \in \N$, we consider the integral
\[ 0 = \int_{\R^3} \bigg| \Big( \frac{\partial}{\partial \xi^0} \Big)^p g(\xi)\Big|_{\xi^0=0} \bigg|^2\: d^3 \xi \:. \]
Applying Plancherel, it follows that
\[ 0 = \int_{\R^3} \big| \hat{g}(\vec{k}) \big|^2\: d^3k \:, \]
where
\[ \hat{g}(\vec{k}) := \int_0^\infty da\:f(a) \int_{-\infty}^\infty d\omega\: \omega^p\:
\delta \big( \omega^2 - |\vec{k}|^2 - a \big)
= \int_0^\infty \big( |\vec{k}|^2+a \big)^{\frac{p-1}{2}}\:f(a)\:da \:. \]
Therefore, the last integral vanishes for almost all~$\vec{k}$ and~$p \in \N$.
Choosing~$p$ odd and taking derivatives with respect to~$\vec{k}$,
we conclude that the distributional integrals
\[ \int_0^\infty \frac{a^p}{(1+a)^q}\:f(a)\:da \]
are well-defined and vanish for all~$p,q \in \N_0$. We choose 
a parameter~$r \in \N$ and a polynomial~${\mathcal{P}}(a)$
which is non-zero except at the origin and the singular points in such a way that the distribution
\beq \label{gadef}
g(a) := \frac{{\mathcal{P}}(a)}{(1+a)^r}\: f(a)
\eeq
is a regular and bounded function and~$\lim_{a \rightarrow 0, \infty} g(a) = 0$.

We want to show that the function~$g$ vanishes. To this end, it is useful to consider
the compactification~$X=[0, \infty]$ as a compact metric space. On this space, we 
introduce the finite measure
\[ d\mu(a) = \frac{da}{(1+a)^2} \:. \]
and regard~$g$ as a vector in the Hilbert space~$L^2(X, d\mu)$.
Assume that~$g$ is non-zero. Using that
the continuous functions on~$X$ are dense in~$L^2(X, d\mu)$,
there is a function~$h \in C^0(X, \R)$ with the property that
\beq \label{notzero}
\int_0^\infty \overline{h}\: g\: d\mu \neq 0 \:.
\eeq
We next consider the algebra generated by the family of functions
\[  h_{p,q}(a) = \frac{a^p}{(1+a)^q} \qquad \text{with~$0 \leq p \leq q$}\:. \]
These functions are all continuous on~$X$ and separate points.
Therefore, according to the Stone-Weierstra{\ss} theorem, there is a sequence~$h_n$
in the algebra which converges uniformly to~$\overline{g}$. Lebesgue's dominated convergence
theorem implies that
\[ 0 = \int_0^\infty \overline{h_n(a)}\: g(a)\: d\mu(a) \longrightarrow 
\int_0^\infty \overline{h}\, g\: d\mu \:, \]
in contradiction to~\eqref{notzero}. This proves that the function~$g$ in~\eqref{gadef} is zero.

We conclude that the distribution~$f$ is supported at a finite number of points,
\[ \supp f \subset \{0, a_1, \ldots, a_N \} \:. \]
If~$f$ is non-zero, there is a polynomial~${\mathcal{P}}$ with the property that~${\mathcal{P}} f$
is a non-trivial multiple of the Dirac distribution supported at one of these singular points. On the other hand,
the distributional integral
\[ \int_0^\infty {\mathcal{P}}(a)\: f(a)\: da \]
vanishes, a contradiction.
\QED

We conclude that causality implies that~$\hat{Q}$ must be of the form
\[ \hat{Q}(p) = \epsilon(p^0)\: \Theta(p^2)\: \big(  f(p^2)\, \slashed{p} + g(p^2) \big) \]
In particular, the solution space of the dynamical wave equation is necessarily symmetric
under reflections~$p \rightarrow -p$. This means that, having solutions on the lower mass shell
of mass~$m$, there must also be solutions on the upper mass shell corresponding to the same mass~$m$.

\Thanks{{{\em{Acknowledgments:}}
We are grateful to the referee for the careful reading and many useful suggestions.

\bibliographystyle{amsplain}
\providecommand{\bysame}{\leavevmode\hbox to3em{\hrulefill}\thinspace}
\providecommand{\MR}{\relax\ifhmode\unskip\space\fi MR }
\providecommand{\MRhref}[2]{%
  \href{http://www.ams.org/mathscinet-getitem?mr=#1}{#2}
}
\providecommand{\href}[2]{#2}

\end{document}